\documentclass[aps,prd,superscriptaddress,groupedaddress,nofootinbib,nobibnotes]{revtex4}

\usepackage{graphicx}      
                                       
\usepackage{dcolumn}
\usepackage{bm}
\usepackage{amssymb}
\usepackage{epstopdf}
\usepackage{amsmath}
\usepackage{amsfonts}
\usepackage{amsthm}
\usepackage{color}
\usepackage{mathrsfs}
\usepackage{hyperref}
\usepackage{dsfont}
\usepackage{empheq}
\usepackage{float}
\usepackage[title]{appendix}
\usepackage[]{breqn}

\usepackage{scalerel}
\usepackage{tikz}
\usetikzlibrary{svg.path}
\definecolor{orcidlogocol}{HTML}{A6CE39}
\tikzset{
  orcidlogo/.pic={
    \fill[orcidlogocol] svg{M256,128c0,70.7-57.3,128-128,128C57.3,256,0,198.7,0,128C0,57.3,57.3,0,128,0C198.7,0,256,57.3,256,128z};
    \fill[white] svg{M86.3,186.2H70.9V79.1h15.4v48.4V186.2z}
                 svg{M108.9,79.1h41.6c39.6,0,57,28.3,57,53.6c0,27.5-21.5,53.6-56.8,53.6h-41.8V79.1z M124.3,172.4h24.5c34.9,0,42.9-26.5,42.9-39.7c0-21.5-13.7-39.7-43.7-39.7h-23.7V172.4z}
                 svg{M88.7,56.8c0,5.5-4.5,10.1-10.1,10.1c-5.6,0-10.1-4.6-10.1-10.1c0-5.6,4.5-10.1,10.1-10.1C84.2,46.7,88.7,51.3,88.7,56.8z};
  }
}
\usepackage{hyperref}  
\hypersetup{colorlinks}

\newcommand\orcidlink[1]{\href{https://orcid.org/#1}{\mbox{\scalerel*{
\begin{tikzpicture}[yscale=-1,transform shape]
\pic{orcidlogo};
\end{tikzpicture}
}{|}}}}

\usepackage{blindtext}
\begingroup\makeatletter
\catcode`,=\active
\global\let\breqn@comma,
\protected\gdef,{\ifmmode\expandafter\breqn@comma\else\expandafter\active@comma\fi}
\endgroup

\newcommand{\be}{\begin{equation}}
\newcommand{\ee}{\end{equation}}
\newcommand{\ba}{\begin{eqnarray}}
\newcommand{\ea}{\end{eqnarray}}

\newcommand{\barr}{\begin{array}}
\newcommand{\earr}{\end{array}}

\newcommand\lsim{\mathrel{\rlap{\lower4pt\hbox{\hskip1pt$\sim$}}
        \raise1pt\hbox{$<$}}}
\newcommand\gsim{\mathrel{\rlap{\lower4pt\hbox{\hskip1pt$\sim$}}
        \raise1pt\hbox{$>$}}}

\begin{document}

\title{Constraining Cosmological Parameters with Needlet Internal Linear Combination Maps II: Likelihood-Free Inference on NILC Power Spectra}
\date{\today}

\author{Kristen M.~Surrao\,\orcidlink{0000-0002-7611-6179}}
\email{k.surrao@columbia.edu}
\affiliation{Department of Physics, Columbia University, New York, NY 10027, USA}

\author{J.~Colin Hill\,\orcidlink{0000-0002-9539-0835}} 
\email{jch2200@columbia.edu}
\affiliation{Department of Physics, Columbia University, New York, NY 10027, USA}

\begin{abstract}
    Standard cosmic microwave background (CMB) analyses constrain cosmological and astrophysical parameters by fitting parametric models to multifrequency power spectra (MFPS). However, such methods do not optimally weight maps in power spectrum (PS) measurements for non-Gaussian CMB foregrounds. We propose needlet internal linear combination (NILC), operating on wavelets with compact support in pixel and harmonic space, as a weighting scheme to yield more optimal parameter constraints. In a companion paper, we derived an analytic formula for NILC map PS, which is physically insightful but computationally difficult to use in parameter inference pipelines. In this work, we analytically show that fitting parametric templates to MFPS and harmonic ILC PS yields identical parameter constraints when the number of sky components equals or exceeds the number of frequency channels. We numerically show that, for Gaussian random fields, the same holds for NILC PS. This suggests that NILC can reduce parameter error bars in the presence of non-Gaussian fields since it uses non-Gaussian information. As Gaussian likelihoods may be inaccurate, we use likelihood-free inference (LFI) with neural posterior estimation. We show that performing inference with auto- and cross-PS of NILC component maps as summary statistics yields smaller parameter error bars than inference with MFPS. For a model with CMB, an amplified thermal Sunyaev--Zel'dovich (tSZ) signal, and noise, we find a 60\% reduction in the area of the 2D 68\% confidence region for component amplitude parameters inferred from NILC PS, as compared to inference from MFPS. Primordial $B$-mode searches are a promising application for our new method, as the amplitude of the non-Gaussian dust foreground is known to be larger than a potential signal. Our code is available in \verb|NILC-Inference-Pipeline|.\footnote{\url{https://github.com/kmsurrao/NILC-Inference-Pipeline}}
\end{abstract}

\maketitle

\section{Introduction} 

\label{sec:intro}

Standard state-of-the-art cosmic microwave background (CMB) data analyses --- such as those used in \textit{Planck} \cite{Planck2018}, the Atacama Cosmology Telescope (ACT) \cite{Dunkley, ACT2020}, the South Pole Telescope (SPT) \cite{SPT}, and BICEP \cite{BICEP:2021xfz} --- fit data to theoretical parametric templates at the power spectrum level using multifrequency data. These analyses have been used to derive the parameter constraints that underlie the current standard model of cosmology. Before computing the frequency-frequency auto- and cross-spectra of the maps, inverse noise variance weighting is applied to the maps to downweight areas of high instrumental noise, allowing one to more optimally recover the signal (e.g., \cite{Atkins:2023yzu, Planck:2015mapcalibration, SPT:2023jql, 2011ApJ...743...90S}). However, this weighting does not downweight CMB foreground contaminants, which are generally non-Gaussian fields. Therefore, the signal that is recovered when applying inverse noise variance weighting is the total sky signal (foregrounds in addition to the CMB). Since the cosmological signal of interest is the CMB alone, we should aim to infer this particular component with the smallest possible error bars. In order to achieve that goal, one should apply a weighting scheme that captures both scale-dependent and spatially varying information to suppress the non-Gaussian CMB foregrounds, improving power spectrum estimation and hence parameter error bars. Since the weighting would be applied at the map level prior to power spectrum estimation, this motivates the exploration of various map-based weighting schemes.

At the map level, there are several methods for constructing component-separated maps of various signals in the microwave sky. Some methods fit parametric models \cite{2006Eriksen, 2008Eriksen}. Others, such as internal linear combination (ILC) \cite{Bennett2003, Tegmark2003, Eriksen2004}, are semi-blind approaches. The ILC procedure estimates a map of some signal of interest by finding the minimum-variance linear combination of the observed frequency maps that satisfies the constraint of unit response to the signal of interest. ILC can be performed in both real/pixel space and harmonic space. 

Additionally, it can be applied on a tight frame of spherical wavelets, known as needlets, in a method called needlet ILC (NILC) \cite{Delabrouille2009}. The benefit of NILC is that needlets have compact support in both pixel and harmonic space, allowing the ILC weights to vary as functions of both angular scale and position. NILC has been used to construct high-resolution maps of individual components in the microwave sky using data or simulations from \emph{Planck}, ACT, SPT, the Simons Observatory, CMB-S4, and \emph{LiteBIRD}, e.g., Refs.~\cite{Coulton2023ACT, McCarthy:2023hpa, Chandran:2023akr, Bleem2022SPT, Remazeilles:2012pn, Wolz:2023lzb, Carones:2022xzs, Abazajian:2019eic}.

The primary idea of this paper is to use NILC as a concrete example of a weighting scheme that optimally downweights contaminants and upweights signal-dominated regions before computing power spectra. In particular, we explore the approach of producing a NILC map for every component in the sky model, computing the auto- and cross-power spectra of the resulting NILC maps, and using these power spectra as summary statistics for parameter inference. We do the same with harmonic ILC (HILC), mathematically proving that it yields the same parameter constraints as the traditional multifrequency power spectrum approach. This suggests that NILC can indeed reduce parameter error bars in the presence of non-Gaussian foregrounds since it uses non-Gaussian information that HILC does not. In a companion paper we developed an analytic formalism by which to parameterize NILC power spectra, which enabled novel insights into the beyond-Gaussian information that the NILC algorithm uses in constructing foreground-cleaned maps~\cite{surrao2024_nilc1}. However, we found that parameter dependence of the NILC component map power spectra was complicated, particularly due to the appearance of terms involving correlations between the NILC weight maps and the underlying fields, and that the analytic results were very slow to use in a realistic computation.

 In this paper, we investigate numerical techniques for parametrizing NILC map power spectra for use in a Gaussian likelihood. However, we also find that the Gaussian likelihood is not a good assumption, and thus use likelihood-free inference (LFI), showing explicitly how discrepant the results from the Gaussian likelihood are. Previous studies have considered the use of likelihood-free inference, also known as simulation-based inference (SBI), to constrain cosmological parameters (e.g.,~\cite{Lemos:2022, Alsing:2019xrx, Wang:2023vej}). LFI can be performed at the field level, which indeed takes into account spatially varying information, but requires large numbers of computationally expensive simulations. It can also be done using summary statistics such as the power spectrum, which allows one to optimally compress the data for Gaussian random fields. Using NILC power spectra as summary statistics allows several advantages as compared to a pure field-level approach. First, a power-spectrum-based approach allows one to use noise-debiased data-split cross-spectra (e.g.,~\cite{WMAP:2003zzr,Planck:2018cosmoparams,ACT_2020_Choi}), mitigating the need for the LFI to learn how to remove instrumental noise. Additionally, nearly all the information for the primary field of interest (the CMB) is contained in its two-point function alone. NILC allows one to better recover that power spectrum by suppressing contributions from non-Gaussian contaminants.

The remainder of this paper is organized as follows. In Sec.~\ref{sec.ILC} we review ILC methods, including harmonic and needlet ILC. In Sec.~\ref{sec.lfi} we review likelihood-free inference (LFI), also known as simulation-based inference (SBI). In Sec.~\ref{sec.methods} we provide an overview of the computational setup and the various methods implemented in this work. In Sec.~\ref{sec.method_multifrequency_power_spec}, \ref{sec.method_hilc}, and \ref{sec.method_nilc}, we describe in detail the various methods and summary statistics considered, using both Gaussian likelihoods and LFI to obtain posteriors. Specifically, we consider the use of multifrequency auto- and cross-power spectra, harmonic ILC map power spectra, and needlet ILC map power spectra as summary statistics, respectively. In Sec.~\ref{sec.results} we present the results of each of the methods, showing how 2D posteriors shrink significantly when using needlet ILC summary statistics. Finally, in Sec.~\ref{sec.discussion} we discuss the results and their implications, particularly for primordial $B$-mode searches. The main results of this work are summarized in Fig.~\ref{fig:final_posteriors} and Table \ref{table:posteriors}, as well as in Fig.~\ref{fig:posteriors_varyfreqs_lfi} and Table \ref{table:ellipse_area} in Appendix \ref{app:3freqs}. Our code is publicly available in \verb|NILC-Inference-Pipeline| and contains pipelines for all the methods described in this work.

\section{Internal Linear Combination}
\label{sec.ILC}

Internal linear combination (ILC) \cite{Bennett2003, Tegmark2003, Eriksen2004, Delabrouille2009} is a method to estimate a map of a signal of interest by finding the minimum-variance linear combination of a set of observed frequency maps that satisfies the constraint of unit response to the signal of interest. Specifically, in pixel space, the signal of interest at some pixel $p$ can be expressed as $\hat{y}(p) = \sum_i w_i \Delta T_i(p)$, where $\Delta T_i(p)$ is the temperature fluctuation in the $i$th frequency map and $w_i$ is the associated linear-combination weight. Then the problem of finding the weights can be expressed as follows: 

\be
\begin{aligned}
\min_{w^i} \quad & \sigma^2_{\hat{y}}=N^{-1}_{\mathrm{pix}} \sum_p (\hat{y}(p)-\langle y \rangle)^2\\
\textrm{s.t.} \quad & \sum_i w^i g_i = 1 \, ,
\end{aligned}
\ee
where $N_{\mathrm{pix}}$ is the number of pixels, $\langle y \rangle$ is the average signal across pixels, and $g_i$ is the spectral response of the component of interest at the $i$th frequency channel. The solution for the weights can be found using Lagrange multipliers \cite{Eriksen2004}:

\be
\label{eq.real_space_cov}
w^i = \frac{g_j(\hat{R}^{-1})_{ij}}{g_k(\hat{R}^{-1})_{kl}g_l} \qquad \text{with} \qquad \hat{R}_{ij}=N^{-1}_{\mathrm{pix}}\sum_p \Delta T_i(p) \Delta T_j(p) \, ,
\ee 
where $\hat{R}_{ij}$ is the empirical frequency-frequency covariance matrix of the observed maps. ILC can also be formulated in harmonic space, giving $\ell$-dependent weights:
\be
\label{eq.HILCweights}
w^i_{\ell} = \frac{ \left( \hat{R}_{\ell}^{-1} \right)_{ij} g_j }{\left( \hat{R}_{\ell}^{-1} \right)_{km} g_k g_m }  \qquad \text{with} \qquad \left(\hat{R}_{\ell} \right)_{ij}= \sum_{\ell' = \ell-\Delta \ell /2}^{\ell+\Delta \ell /2} \frac{2\ell'+1}{4\pi} C_{\ell'}^{ij} \, .
\ee 
The multipole bin width $\Delta \ell$ (or the pixel domain size in real-space ILC above) must be large enough to mitigate the ``ILC bias" that results from computing the covariances for ILC weights using a small number of modes \cite{Delabrouille2009}. 

To maximize the robustness of the ILC procedure, it is frequently applied to CMB data on a needlet frame \cite{Delabrouille2009, Planck:2013compsep, Planck:2013ymap, Planck2015compsep, Planck2015ymap, Planck2018, Chandran:2023akr, Coulton2023ACT, McCarthy:2023cwg, McCarthy:2023hpa}. Needlets are a set of basis functions on the sphere that possess compact support in both real space and harmonic space \cite{Narcowich2006, Marinucci_2007}, allowing one to obtain ILC weights that vary both as a function of scale (depending on multipole $\ell$) and of position (depending on direction or spatial pixel $\hat{n}$). Such a procedure allows us to apply an optimal weighting scheme to non-Gaussian and/or anisotropic foregrounds.  We briefly summarize the NILC procedure below \cite{Guilloux2008, Delabrouille2009}.

Consider a set of frequency maps $T^i(\hat{n})$, where $i$ is an index denoting the frequency channel of each map in our data set. For a needlet filter $h^{(n)}_{\ell}$, indexed by $(n)$ ranging from $1$ to $N_{\rm scales}$ (the total number of needlet filters), the NILC operations on this set of frequency maps are as follows:
\begin{itemize}
\item Transform each frequency map to harmonic space and filter it with the needlet filter function $h^{(n)}_{\ell}$:
\begin{equation}
\label{eq.filterstep1}
T^i_{\ell m} \rightarrow T^{i,(n)}_{\ell m} \equiv T^i_{\ell m} h^{(n)}_{\ell} \,.
\end{equation}
This procedure produces $N_{\rm scales}$ separate maps for each frequency channel (each frequency map gets filtered by each needlet filter separately).

\item Define local pixel-space domains, and compute the smoothed frequency-frequency covariance matrix on each domain. In particular, let $\mathcal{D}^{(n)}_{\alpha}$ denote a real-space domain in frequency maps that have been filtered with needlet scale $(n)$, where $\alpha$ labels each domain on that map. The frequency-frequency covariance matrix is then 
\begin{equation}
    \label{eq.NILCcov}
    (\hat{R}_\alpha^{(n)})_{ij} = N_{\mathrm{pix}}^{-1} \sum_{p \in \mathcal{D}^{(n)}_{\alpha}} \Delta T_i(p) \Delta T_j(p) \, ,
\end{equation}
where $N_{\mathrm{pix}}$ is the number of pixels in $\mathcal{D}^{(n)}_{\alpha}$. This equation is nearly identical to Eq.~\eqref{eq.real_space_cov} for the real-space ILC frequency-frequency covariance matrix, except that here the covariance matrix is computed independently for each needlet filter scale. In practice, Eq.~\eqref{eq.NILCcov} is usually implemented by smoothing the product map $\Delta T_i \Delta T_j$ with a Gaussian kernel, with larger kernels used for lower-multipole needlet filter scales.

\item For each frequency channel and each needlet filter scale, determine a map of weights in pixel space $W^{i,(n)}(\hat{n})$ analogously to Eq.~\eqref{eq.real_space_cov}. The weight maps are determined via the ILC algorithm, similar to Eq.~\eqref{eq.real_space_cov} but performed on the local pixel-space domains from the previous step. Multiply each filtered frequency map by the associated weight map:
\begin{equation}
\label{eq.ILCweightingstep}
T^{i,(n)}(\hat{n}) \rightarrow \tilde{T}^{i,(n)}(\hat{n}) \equiv T^{i,(n)}(\hat{n}) W^{i,(n)}(\hat{n})
\end{equation}

\item Add up the ILC-weighted maps to obtain a single ILC map at each needlet filter scale:
\begin{equation}
\label{eq.ILCadditionstep}
T^{{\rm NILC},(n)}(\hat{n}) = \sum_i \tilde{T}^{i,(n)}(\hat{n})
\end{equation}

\item Apply the needlet filter $h^{(n)}_{\ell}$ again:
\begin{equation}
\label{eq.filterstep2}
T^{{\rm NILC},(n)}_{\ell m} \rightarrow T^{{\rm NILC},(n),(n)}_{\ell m} \equiv T^{{\rm NILC},(n)}_{\ell m} h_{\ell}^{(n)}
\end{equation}

\item Sum the results from all needlet filter scales to obtain the final NILC map:
\begin{equation}
\label{eq.ILCfinal}
T^{\rm NILC}(\hat{n}) = \sum_{(n)} T^{{\rm NILC},(n),(n)} (\hat{n})
\end{equation}

\end{itemize}

Producing a NILC map in this manner allows the signal of interest to propagate in an unbiased fashion to the final map, due to the ILC signal-preservation constraint and the NILC filter power-preservation constraint, $\sum_{(n)} \left( h_{\ell}^{(n)} \right)^2 = 1$ at each $\ell$. However, contaminant signals propagate in a nontrivial way, as derived in Ref.~\cite{surrao2024_nilc1}. 

Thus far, we we have described ILC methods that remove contaminants by minimizing the variance of the final map. It is also possible to perform constrained ILC methods, which explicitly deproject contaminants by imposing an additional constraint that the response of the final map to the contaminant must be zero \cite{2009Chen, 2011Remazeilles}. However, the additional constraint comes at the cost of increasing the variance of the final map. In this work, we consider only standard ILC methods (no explicit deprojection of contaminants) throughout. 

\section{Likelihood-Free Inference}
\label{sec.lfi}

In standard Bayesian analysis, given some parameters $\theta$ and an observation $x$, one can compute the posterior $p(\theta | x)$ using
\begin{equation}
    p(\theta | x) = \frac{p(x|\theta)p(\theta)}{\int d\theta^{\prime} p(x|\theta^{\prime})p(\theta^{\prime})} \,,
\end{equation}
where $p(x|\theta)$ is the likelihood, $p(\theta)$ is the prior, and $p(x)=\int d\theta^{\prime} p(x|\theta^{\prime})p(\theta^{\prime})$ is the evidence. 

In certain situations, however, it is intractable to construct an analytic likelihood. Likelihood-free inference (LFI), also known as simulation-based inference (SBI) or implicit likelihood inference, is a method for determining posterior distributions without using an explicit likelihood (e.g., \cite{2020PNAS..11730055C}). In particular, given some prior $p(\theta)$, LFI requires a simulator that samples from the prior and generates simulations, i.e., $x_i \sim p(x_i | \theta_i)$. With several simulations on hand, one can then train a normalizing flow network to learn the likelihood (``neural likelihood estimation") \cite{SNLE1, SNLE2, SNLE3}, to learn the likelihood-evidence ratio (``neural ratio estimation") \cite{SNRE1, SNRE2, SNRE3, SNRE4}, or to learn the posterior directly (``neural posterior estimation") \cite{SNPE1, SNPE2, SNPE3, SNPE4}. When the length of the data vector being simulated is much larger than the number of parameters, as is the case in our set-up discussed in the coming sections, neural posterior estimation (NPE) generally performs best.

\subsection{Masked Autoregressive Flows}

There are various possible choices for the exact network that is used in LFI. In our work, we use masked autoregressive flows (MAF) \cite{MAF}, for which we find the most stable training results. MAF combines the advantages of autoregressive models \cite{autoregressive_model}, which model several conditionals whose product comprises the target joint density, and normalizing flows \cite{normalizing_flow}, which use invertible transformations to convert a base density into the target density. In the remainder of this section, we follow the notation of Ref.~\cite{MAF}.

Autoregressive models decompose the target density $p(\mathbf{x})$ as 
\begin{equation}
    p(\mathbf{x}) = \prod_i p(x_i |\mathbf{x}_{1:i-1}),
\end{equation}
where the subscript $1:i-1$ denotes that the $i$th state is dependent on all previous states indexed from 1 to $i-1$. The parameters of each conditional can be learned as functions of hidden states of recurrent networks. 

Normalizing flows take some base density $\pi_u (\mathbf{u})$, draw $\mathbf{u} \sim \pi_u (\mathbf{u})$, and let $\mathbf{x} = f(\mathbf{u})$, where $f$ is an invertible transformation. The joint target density $p(\mathbf{x})$ is computed as
\begin{equation}
    p(\mathbf{x}) = \pi_u(f^{-1}(\mathbf{x})) \left | \mathrm{det} \left ( \frac{\partial f^{-1}}{\partial \mathbf{x}} \right ) \right | \, ,
\end{equation}
and thus, $f$ must have a tractable Jacobian.

Autoregressive models can be treated as normalizing flows: Let 
\begin{equation}
    p(x_i | \mathbf{x}_{1:i-1}) = \mathcal{N}(x_i | \mu_i, (\mathrm{exp} \, \alpha_i)^2)
\end{equation}
with $\mu_i = f_{\mu_i}(\mathbf{x}_{1:i-1})$ and $\alpha_i = f_{\alpha_i}(\mathbf{x}_{1:i-1})$, where $f_{\mu_i}$ and $f_{\alpha_i}$ are scalar functions. Then taking some random vector $\mathbf{u}$ with $u_i \sim \mathcal{N}(0,1)$, $x_i=u_i \, \mathrm{exp} \, \alpha_i + \mu_i$. This allows one to start with some random seed and transform it into the target density.

MAF is made by stacking several of the Masked Autoencoder(s) for Distribution Estimation (MADE) \cite{MADE}. MADE uses a feedforward network to learn $f_{\mu_i}$ and $f_{\alpha_i}$, applying binary masks to the weight matrices to enforce the autoregressive property. We refer the reader to Ref.~\cite{MAF} for more details.

\section{Overview of Methods and Computational Setup}
\label{sec.methods}

\subsection{General Computational Setup}
As a simple, demonstrative example, we consider a simulated sky model comprising two components: the CMB signal and the thermal Sunyaev-Zel'dovich (tSZ) signal, which is amplified by a factor of 150 at the map level (for reasons described below). The tSZ effect is the inverse-Compton scattering of CMB photons off hot electrons and is a useful probe of hot gas in galaxy clusters \cite{SZ1969, SZ1970}. We refer to the amplified tSZ signal as the ``fake'' tSZ or ``ftSZ'' field. In CMB thermodynamic temperature units, the CMB spectral response is unity at every frequency. We use units of K for the CMB field. The tSZ spectral response, which we denote as $g(\nu)$ with $\nu$ being some frequency in Hz, is given by 
\begin{equation}
    \label{eq.tsz_sed}
    g(\nu) = x \coth{\frac{x}{2}} - 4 \; \text{ with } \; x=\frac{h\nu}{k_B T_{\mathrm{CMB}}} \, ,
\end{equation}
where $h$ is Planck's constant, $k_B$ is the Boltzmann constant, and $T_{\mathrm{CMB}}=2.726$ is the CMB temperature today (at redshift $z=0$).

We simulate these sky components at two frequencies, 90 and 150 GHz. We then add Gaussian, white noise to the mock sky map for each frequency channel.  Lensed $a_{\ell m}$ for the CMB component are obtained from the WebSky Extragalactic CMB Mocks\footnote{\url{https://mocks.cita.utoronto.ca/index.php/WebSky_Extragalactic_CMB_Mocks}}, which assume a flat $\Lambda$CDM cosmology with cosmological parameters consistent with Planck 2018 \cite{Planck:2018cosmoparams}: ($\Omega_m$, $\Omega_b$, $\sigma_8$, $n_s$, $h$, $\tau$) = (0.31, 0.049, 0.81, 0.965, 0.68, 0.055) \cite{Websky_2020}. The \verb|HEALPix/healpy| \cite{Healpix, Healpy} software package is used to compute some fiducial power spectrum and then generate many CMB realizations from that spectrum. For the tSZ maps, we use \verb|halosky|\footnote{\url{https://github.com/marcelo-alvarez/halosky}}, which generates realistic random non-Gaussian tSZ simulations from random halo catalogs. Specifically, \verb|halosky| Poisson samples from the Tinker et al.~(2008) halo mass function \cite{Tinker:2008} (determining halo abundance by using the WebSky \cite{Websky_2020} linear matter power spectrum based on the \emph{Planck} 2018 cosmological parameters \cite{Planck:2018cosmoparams}) and then populates a catalog of halos along the lightcone.  The Battaglia et al.~(2012) AGN feedback pressure profile is used \cite{Battaglia_2012}. We use redshift limits $0 \leq z \leq 5.0$ and mass limits $5 \times 10^{14} \,\, M_{\odot} \leq M \leq 10^{16} \,\, M_{\odot}$ in this construction, the latter chosen for computational efficiency. This narrow mass range is not important since we are rescaling the tSZ field by a large amplitude in this toy model. It also has the effect of populating fewer halos, making the tSZ effect more Poissonian and thereby more non-Gaussian. This is useful for our demonstration purposes, the reasoning for which is discussed further below. 

\begin{figure}[t]
    \centering
    \includegraphics[width=0.55\textwidth]{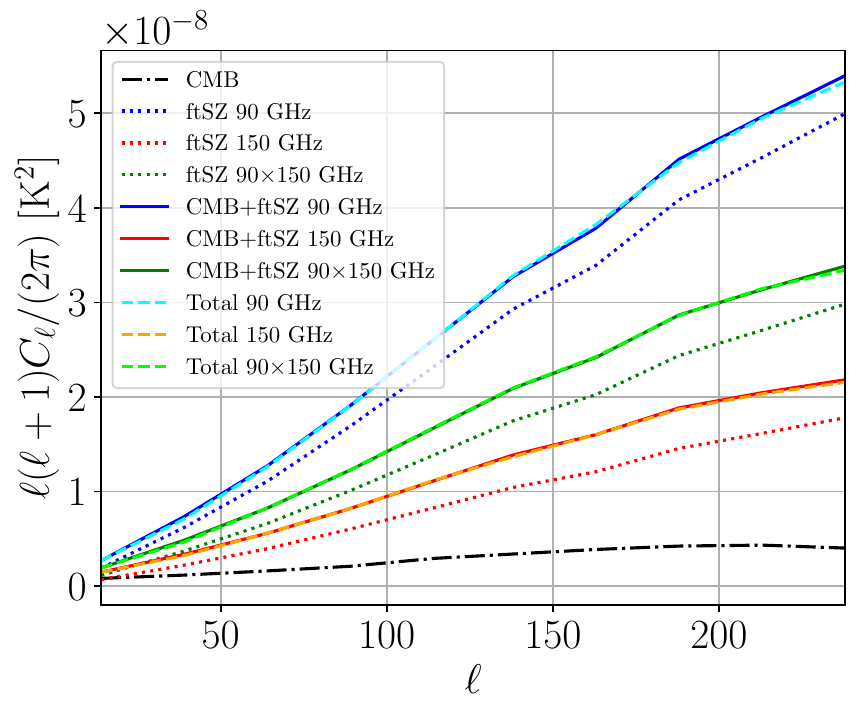}
    \caption{Sample simulated power spectra (in units of $\mathrm{K}^2$), plotted as $D_\ell = \ell(\ell+1)C_\ell/(2\pi)$. Plotted are the CMB power spectrum (black dash-dotted curve); ``ftSZ" or amplified tSZ auto-spectrum at 90 GHz (dotted blue curve), auto-spectrum at 150 GHz (dotted red curve), and 90 $\times$ 150 GHz cross-spectrum (dotted green curve). The combined CMB+ftSZ spectra at 90, 150, and 90 $\times$ 150 GHz are also shown (solid blue, solid red, and solid green curves, respectively). Finally, the total power spectra at 90, 150, and 90 $\times$ 150 GHz are shown (dashed cyan, dashed orange, and dashed lime curves, respectively). Of note is that there is no noise bias in the total power spectrum since we have used noise-debiased data-split cross-spectra, though there are noise fluctuations.}
    \label{fig:input_spectra}
\end{figure}

To avoid noise biases in the measured power spectra, we use data-split cross-spectra, as is done in many actual CMB analyses (e.g.,~\cite{WMAP:2003zzr,Planck:2018cosmoparams,ACT_2020_Choi}). For the noise power spectrum in each simulated split map, we use the model given by Ref.~\cite{Knox1995}:
\begin{equation*}
    \label{eq.Nell}
    N_\ell = W^2 e^{\ell(\ell+1)\sigma^2} \qquad \text{with} \qquad \sigma = \theta_{\rm FWHM} / \sqrt{8 \ln 2}\,,
\end{equation*}
where we set $W_{90} = W_{150} = 3 \times 10^4 \, \mu {\rm K}_{\rm CMB} \cdot {\rm arcmin}$  for each noise split and $\theta_{\rm FWHM} = 1.4$ arcmin for both the 90 and 150 GHz beams in our simulations. The large noise is important for our simple two-frequency, two-component-plus-noise sky model, where we do not want the NILC maps to completely clean the contaminating foregrounds (thus giving a realistic representation on what would happen with actual data where there are several sky components that cannot all be simultaneously cleaned completely). All maps are generated at HEALPix resolution parameter $N_{\mathrm{side}}=128$, and power spectra are computed up to $\ell_{\mathrm{max}}=250$ for computational efficiency on this simple, demonstrative example. We then add the CMB, ftSZ, and noise map at each frequency and each split, resulting in four maps (two frequencies and two splits). An example of the simulated component power spectra and frequency-frequency power spectra is shown in Fig.~\ref{fig:input_spectra}.

Our goal is to constrain two parameters, $A_{\mathrm{CMB}}$ and $A_{\mathrm{ftSZ}}$, which are two overall amplitude parameters that uniformly scale the CMB and ftSZ power spectrum over the entire multipole range, respectively. The fiducial parameter values are $A_{\mathrm{CMB}} = 1 = A_{\mathrm{ftSZ}}$. Specifically,
\begin{equation}
    C_{\ell}^{ij}(A_{\rm CMB}, A_{\rm ftSZ}) = A_{\rm CMB} C_{\ell}^{\mathrm{CMB}} + A_{\rm ftSZ} g^i_y g^j_y C_{\ell}^{\mathrm{ftSZ}} \, ,
\end{equation}
where $i$ and $j$ index frequency channels, $C_\ell^{ij}$ is the frequency-frequency auto- or cross-power spectrum, $C_\ell^{\mathrm{CMB}}$ is a CMB power spectrum template, $C_\ell^{\mathrm{ftSZ}}$ is a power spectrum template of the amplified Compton-$y$ field, and $g^i_y$ and $g^j_y$ denote the tSZ spectral responses at frequencies $i$ and $j$. We could have also defined the amplitude parameters at the  map level, but we choose to define them at the power spectrum level to be consistent with what is done in current major CMB data analyses (and also because the primary field of interest, the CMB, is entirely described by its power spectrum). We consider various methods for obtaining the posteriors on these parameters, outlined in the next subsection.

The tSZ field is highly non-Gaussian, as it traces hot gas in galaxy clusters. At the low multipole values considered in this work, the tSZ power spectrum lies significantly below that of the CMB. As described in Sec.~\ref{sec:intro}, our goal is to show that NILC decreases parameter error bars in the presence of non-Gaussian fields. To demonstrate that idea, we would need the tSZ power spectrum to be comparable to or larger than the CMB power spectrum in magnitude. Thus, each tSZ map is amplified by a factor of 150 for our simulations, since we are interested in showing that NILC-based summary statistics are an improvement over multifrequency power spectra summary statistics for data with large non-Gaussian components. 


\subsection{Outline of Implemented Methods}
In this subsection we provide an outline of the methods considered in this work. We describe each method in detail in Sec.~\ref{sec.method_multifrequency_power_spec}, \ref{sec.method_hilc}, and \ref{sec.method_nilc}.

\begin{enumerate}
    \item Multifrequency Power Spectra (see Sec.~\ref{sec.method_multifrequency_power_spec})
        \begin{enumerate}
            \item Gaussian Likelihood (with known analytic parameter dependence)
                \begin{enumerate}
                    \item Maximum Likelihood Estimation (MLE)
                    \item Fisher Matrix
                    \item Markov chain Monte Carlo (MCMC) Algorithm
                \end{enumerate}
            \item Likelihood-Free Inference (LFI)
        \end{enumerate}

    \item Harmonic ILC (HILC) Power Spectra (see Sec.~\ref{sec.method_hilc})
        \begin{enumerate}
            \item Gaussian Likelihood
                \begin{itemize}
                    \item Fixed weights
                        \begin{itemize}
                            \item Using known analytic parameter dependence
                                \begin{enumerate}
                                    \item MLE
                                    \item Fisher Matrix
                                    \item MCMC Algorithm
                                \end{enumerate}
                            \item Parameter dependence determined via symbolic regression (SR)
                                \begin{enumerate}
                                    \item MLE
                                    \item Fisher Matrix
                                    \item MCMC Algorithm
                                \end{enumerate}
                        \end{itemize}
                    
                    \item Varying weights
                        \begin{itemize}
                            \item Parameter dependence determined via SR
                                 \begin{enumerate}
                                    \item MLE
                                    \item Fisher Matrix
                                    \item MCMC Algorithm
                                \end{enumerate}
                        \end{itemize}
                \end{itemize}
            \item LFI
                \begin{itemize}
                    \item Fixed weights
                    \item Varying weights
                \end{itemize}
        \end{enumerate}

    \item Needlet ILC (NILC) Power Spectra (see Sec.~\ref{sec.method_nilc})
        \begin{enumerate}
            \item Gaussian Likelihood (with parameter dependence determined via SR)
                 \begin{enumerate}
                    \item MLE
                    \item Fisher Matrix
                    \item MCMC Algorithm
                \end{enumerate}
            \item LFI 
        \end{enumerate}
\end{enumerate}

\section{Multifrequency Power Spectra as Summary Statistics}
\label{sec.method_multifrequency_power_spec}

\subsection{Gaussian Likelihood}
\label{sec.method_multifrequency_power_spec_gaussian_lkl}
Fitting multifrequency auto- and cross-power spectrum data to parametric templates using a Gaussian likelihood is the classic method used in nearly all major primary CMB analyses to date. In our case, we use the auto- and cross-spectra at the two frequency channels as our data vector. Specifically, our data vector consists of $C_b^{90 \mathrm{ split1}, 90 \mathrm{split2}}$, $C_b^{90 \mathrm{ split1}, 150 \mathrm{split2}}$, $C_b^{150 \mathrm{ split1}, 90 \mathrm{split2}}$, and $C_b^{150 \mathrm{ split1}, 150 \mathrm{split2}}$, where $90\mathrm{split1}$ signifies the 90 GHz map with the first noise split, and similarly for the other frequency-split pairs. The subscript $b$ denotes the multipole $\ell$-space bin, where we use 10 linearly spaced bins from $\ell=2$ to $\ell=250$ (one bin contains 24 multipoles, and all other bins contain 25 multipoles). The Gaussian likelihood is given by 

\begin{align}
    \label{eq.multifrequency_lkl}
    \ln \mathcal{L}(A_{\rm CMB}, A_{\rm ftSZ}) &= -\frac{1}{2} \left( C_{b_1}^{ij}(A_{\rm CMB}, A_{\rm ftSZ}) - C_{b_1}^{ij, \mathrm{data}} \right)  \left(\mathrm{Cov}_{ij b_1,kl b_2}\right)^{-1} \left( C_{b_2}^{kl}(A_{\rm CMB}, A_{\rm ftSZ}) - C_{b_2}^{kl, \mathrm{data}} \right) \, ,
\end{align}
where there is implied summation over repeated indices $i$, $j$, $k$, $l$, $b_1$, and $b_2$ on the RHS of Eq.~\eqref{eq.multifrequency_lkl}. Here $C_{b_1}^{ij}, C_{b_2}^{kl} \in \{ C_b^{90 \mathrm{ split1}, 90 \mathrm{split2}}, C_b^{90 \mathrm{ split1}, 150 \mathrm{split2}}, C_b^{150 \mathrm{ split1}, 90 \mathrm{split2}},  C_b^{150 \mathrm{ split1}, 150 \mathrm{split2}} \}$ and $\mathrm{Cov}_{ij b_1,kl b_2} \equiv \mathrm{Cov}(C_{b_1}^{ij}, C_{b_2}^{kl})$ is the covariance matrix of the multifrequency power spectra, computed directly from 2000 independent simulations generated with the fiducial parameter values (i.e., there is no assumption of a Gaussian covariance matrix). We multiply the inverse of the multifrequency power spectrum covariance matrix by the correction factor $\frac{n-p-2}{n-1}$, where $n$ is the number of simulations and $p$ is the length of the data vector \cite{Hartlap:2006}. Since the number of simulations is far greater than the length of the data vector, this factor is very close to 1 in our case.

Note also that we have included off-diagonal $\ell$-bins in the covariance matrix and likelihood. For the multifrequency power spectra, 
\begin{equation}
    \label{eq.multifrequency_data_vec}
    C_{b_1}^{ij}(A_{\rm CMB}, A_{\rm ftSZ}) = A_{\rm CMB} C_{b_1}^{\mathrm{CMB}} + A_{\rm ftSZ} g^i_y g^j_y C_{b_1}^{\mathrm{ftSZ}} \, ,
\end{equation}
where $g^i_y$ and $g^j_y$ are the spectral responses of the tSZ signal at frequencies $i$ and $j$ (as defined in Eq.~\ref{eq.tsz_sed}), $C_{b_1}^{\mathrm{CMB}}$ is the fiducial CMB power spectrum template, and $C_{b_1}^{\mathrm{ftSZ}}$ is the fiducial (amplified) Compton-$y$ power spectrum template. For these fiducial templates, we use the mean of the CMB and ftSZ spectra generated from 2000 simulations to obtain smooth theory curves. We implement three methods for computing the posteriors on $A_{\mathrm{CMB}}$ and $A_{\mathrm{ftSZ}}$: maximum likelihood estimation (MLE), the Fisher information matrix, and an MCMC algorithm. 

For MLE, we numerically find the maximum-likelihood point for the likelihood function in Eq.~\eqref{eq.multifrequency_lkl} for each of the 2000 simulations. Specifically, for each simulation, we compute the power spectra of the frequency maps to use as $C_{b_1}^{ij,\mathrm{data}}$ and $C_{b_2}^{kl,\mathrm{data}}$. Note that these are noise-debiased by construction, though there are noise fluctuations present. The total power spectrum at each frequency, along with the contributions from individual components, can be seen in Fig.~\ref{fig:input_spectra}. We then find the maximum-likelihood $(A_{\mathrm{CMB}}, A_{\mathrm{ftSZ}})$ point for every simulation using Nelder-Mead minimization \cite{Nelder-Mead} of the negative log likelihood. To avoid getting stuck in local minima, we set run the minimization routine using three starting points: $(A_{\mathrm{CMB}}, A_{\mathrm{ftSZ}}) = (1.0, 1.0)$, $(0.5, 0.5)$, and $(1.5, 1.5)$. We then use the global minimum from these results as our maximum-likelihood point $(\hat{A}_{\mathrm{CMB}}, \hat{A}_{\mathrm{ftSZ}})$. The majority of the time, the results from each of these starting points are equal, and local minima do not pose a problem. Only in a few simulations are the results actually different depending on starting point.

For the Fisher matrix calculation, we obtain the parameter covariance matrix using the usual 
\be
    \sigma^2_{\hat{A}_{\alpha} \hat{A}_{\beta}} = \left( F_{\hat{A}_{\alpha} \hat{A}_{\beta}} \right)^{-1} \, ,
\ee
where $\hat{A}_{\alpha}, \hat{A}_{\beta} \in \{ \hat{A}_{\mathrm{CMB}}, \hat{A}_{\mathrm{ftSZ}} \}$ and $F_{\hat{A}_{\alpha} \hat{A}_{\beta}}$ is the Fisher matrix defined as
\be
\label{eq.Fisher_mult}
    F_{\hat{A}_{\alpha} \hat{A}_{\beta}} = 
    - \left. \left( \frac{ \partial^2 \ln{\mathcal{L}} }{ \partial A_{\alpha} A_{\beta} } \right) \right|_{\hat{A}_{\alpha} \hat{A}_{\beta}}
    =  \left. \frac{\partial C_{b_1}^{ij}}{\partial A_{\alpha}} (\mathrm{Cov}^{-1}_{ijb_1,klb_2}) \frac{\partial C_{b_2}^{kl}}{\partial A_{\beta}}   \right|_{A_{\alpha}=A_{\beta}=1} \, ,
\ee
where there is implied summation over $i$, $j$, $k$, $l$, $b_1$, and $b_2$. As seen from Eq.~\eqref{eq.multifrequency_data_vec}, $\partial C_{b_1}^{ij}/ \partial A_{\mathrm{CMB}}=C_{b_1}^{\mathrm{CMB}}$ and $\partial C_{b_1}^{ij}/ \partial A_{\mathrm{ftSZ}}=C_{b_1}^{\mathrm{ftSZ}}$. 

Finally, for the MCMC algorithm we use \verb|emcee| \cite{emcee} with the likelihood in Eq.~\eqref{eq.multifrequency_lkl} and an observation consisting of the mean of multifrequency auto- and cross-spectra over all simulations. We use ten walkers, each with a random starting point between $(A_{\mathrm{CMB}}, A_{\mathrm{ftSZ}}) = (0.8, 0.8)$ and $(A_{\mathrm{CMB}}, A_{\mathrm{ftSZ}}) = (1.2, 1.2)$. The autocorrelation time is used to assess MCMC convergence. Specifically, the chain is marked as converged when it is longer than 50 times the integrated autocorrelation time. 

\subsection{Likelihood-Free Inference}
\label{sec.method_multifrequency_power_spec_lfi}
For LFI/SBI, we use the code \verb|sbi| \cite{sbi}. As discussed in Sec.~\ref{sec.lfi}, we require four components for LFI: a prior, a simulator, an observation, and a network. For the former, we set a uniform prior centered on 1 with lower bound $(1-5\sigma_{\hat{A}_{\mathrm{CMB}}}, 1-5\sigma_{\hat{A}_{\mathrm{ftSZ}}})$ and upper bound $(1+5\sigma_{\hat{A}_{\mathrm{CMB}}}, 1+5\sigma_{\hat{A}_{\mathrm{ftSZ}}})$, where $\sigma_{\hat{A}_{\mathrm{CMB}}}$ and $\sigma_{\hat{A}_{\mathrm{ftSZ}}}$ are the 1D marginalized errors obtained from the multifrequency Gaussian likelihood on $\hat{A}_{\mathrm{CMB}}$ and $\hat{A}_{\mathrm{tSZ}}$, respectively. The $5\sigma$ range for the prior ensures that the prior is narrow enough to be efficient in the number of simulations, but wide enough to encompass the true error on the parameters ($5\sigma$ of the results from the Gaussian likelihood should be sufficient since the Gaussian likelihood, though not entirely accurate, should not be a catastrophically bad approximation due to the central limit theorem). As a further check that the prior is wide enough, we ensure that the final posteriors do not hit the prior edges.  

For the simulator, we draw $A_{\mathrm{CMB}}$ and $A_{\mathrm{ftSZ}}$ from the prior distribution. Then for each simulation, we multiply the input CMB map by a factor of $\sqrt{A_{\mathrm{CMB}}}$ and the ftSZ map by a factor of $\sqrt{A_{\mathrm{ftSZ}}}$ (recall that the parameters are defined at the power spectrum level). We then proceed as usual with constructing maps of two noise splits for each frequency and constructing the data vector comprising the concatenation of $C_b^{90 \mathrm{ split1}, 90 \mathrm{split2}}$, $C_b^{90 \mathrm{ split1}, 150 \mathrm{split2}}$, $C_b^{150 \mathrm{ split1}, 90 \mathrm{split2}}$, and $C_b^{150 \mathrm{ split1}, 150 \mathrm{split2}}$. Thus, the data vector has a total length of 40 since there are 10 bins in each of the concatenated spectra. We generate 40000 simulations with parameters drawn randomly from the prior. For the observation vector, we generate 2000 simulations of the data vector with $A_{\mathrm{CMB}}=A_{\mathrm{ftSZ}}=1$ and take the mean over the 2000 simulations. This may result in a slightly biased central value of the posteriors since we use significantly fewer simulations to determine the mock observation vector than to determine the full posteriors. However, the focus of this paper is on the spread in the posteriors rather than central values. Note also that the observation vector here is allowed to be smooth since the simulator generates noisy spectra from which the LFI learns the covariance.

We use single-round neural posterior estimation (NPE) for the LFI. We apply z-score normalization to each simulated data vector and to the observation for more stable training. For the network, we train a masked autoregressive flow (MAF). We also experiment with using a neural spine flow, mixture density network, and masked autoencoder for distribution estimation, but we find that the MAF is the most efficient in the number of simulations for our purposes. We use 90\% of the 40000 simulations for training and 10\% for validation. We tune the hyperparameters of the network with the Weights \& Biases framework (\verb|wandb|) \cite{wandb}. Our \verb|wandb| project is public.\footnote{\url{https://wandb.ai/kmsurrao/cmb_sbi/sweeps}} We use a random hyperparameter search with 40 trials. For the final posteriors, we use an ensemble of the top 10 networks in terms of logarithmic probability on the validation set and uniformly weight each of the top 10 networks to obtain the final posteriors. To assess whether reductions in the size of the posterior are significant, we examine the variance in the final parameter covariance matrices obtained from different hyperparameter settings (the ``error on the error"). We find that this ``error on the error" is not significant when using just the top 10 networks. However, the final posteriors are still sensitive to the hyperparameter tuning, as some networks cause the posteriors to hit the prior edges and are thus essentially not learning any information. A table of hyperparameters that are tuned is shown in Table~\ref{tab:hyperparameters}. Specifically, we tune the learning rate, the number of transforms in the normalizing flow, the number of hidden features, the value at which to clip the total gradient norm in order to prevent exploding gradients, and the number of epochs to wait for improvement on the validation set before terminating training.

\begin{table}[t]
    \setlength{\tabcolsep}{10pt}
    \renewcommand{\arraystretch}{1.3}
    \centering
    \begin{tabular}{|c|c|c|c|}
        \hline
         \textbf{Hyperparameter} & \textbf{Minimum Value} & \textbf{Maximum Value} & \textbf{Distribution} \\
         \hline 
         Learning Rate & $1.0 \times 10^{-4}$ & $7.0 \times 10^{-4}$ &uniform \\
         \hline
         Number of Transforms & 3 &10 &integer uniform  \\
          \hline
        Hidden Features &35 &65 &integer uniform  \\
          \hline
        Clip Maximum Norm &3.0 &7.0 &uniform  \\
          \hline
        Epochs for Convergence Assessment &20 &60&integer uniform  \\
          \hline
    \end{tabular}
    \caption{Distributions of hyperparameters tuned in the MAF used for LFI. ``Clip maximum norm'' refers to the value at which to clip the total gradient norm in order to prevent exploding gradients, and ``epochs for convergence assessment'' the number of epochs to wait for improvement on the validation set before terminating training. See Sec.~\ref{sec.method_multifrequency_power_spec_lfi} for details of how the hyperparameter sweep is used. The same distributions of hyperparameters are used to tune the networks used for LFI with multifrequency power spectra, harmonic ILC power spectra, and needlet ILC power spectra summary statistics, described in Sec.~\ref{sec.method_multifrequency_power_spec_lfi}, ~\ref{sec.method_hilc_lfi}, and ~\ref{sec.method_nilc_lfi}, respectively.}
    \label{tab:hyperparameters}
\end{table}

\section{Harmonic ILC Component Map Power Spectra as Summary Statistics}
\label{sec.method_hilc}

Our data vector for the harmonic ILC power spectra consists of $C_b^{\hat{T} \mathrm{ split1}, \hat{T} \mathrm{split2}}$, $C_b^{\hat{T} \mathrm{ split1}, \hat{y} \mathrm{split2}}$, $C_b^{\hat{y} \mathrm{ split1}, \hat{T} \mathrm{split2}}$, and $C_b^{\hat{y} \mathrm{ split1}, \hat{y}\mathrm{split2}}$, where $\hat{T}\mathrm{split1}$ signifies a HILC CMB map produced using the frequency maps with the first noise split, $\hat{y}\mathrm{split1}$ signifies a HILC Compton-$y$ map produced using the frequency maps with the first noise split, and similarly for the second split. The subscript $b$ denotes the multipole $\ell$-space bin, where we use 10 linearly spaced bins from $\ell=2$ to $\ell=250$ (one bin contains 24 multipoles, and all other bins contain 25 multipoles).

We compute the HILC weights and power spectra analytically. As an example, to compute $C_b^{\hat{T} \mathrm{ split1}, \hat{y} \mathrm{split2}}$ we do the following. First, we produce four maps: a 90 GHz map with the first noise split, a 90 GHz map with the second noise split, a 150 GHz map with the first noise split, and a 150 GHz map with the second noise split. We then compute weights for the HILC CMB map using the first noise split in an analogous way to Eq.~\eqref{eq.HILCweights} (and noting the the CMB SED is a vector of ones as described in Sec.~\ref{sec.methods}): 
\begin{equation}
    w^{i,\hat{T}\mathrm{split1}}_\ell = \frac{ \mathds{1}_j \left( \hat{R}_{\ell}^{-1} \right)_{ij} }{ \left( \hat{R}_{\ell}^{-1} \right)_{km} \mathds{1}_k \mathds{1}_m}
    \qquad \text{with} \qquad 
    \left(\hat{R}_{\ell} \right)_{ij}= \sum_{\ell' = \ell-\Delta \ell /2}^{\ell+\Delta \ell /2} \frac{2\ell'+1}{4\pi} C_{\ell'}^{i\mathrm{split1},j\mathrm{split1}} \, ,
\end{equation}
where $i\mathrm{split1}, j\mathrm{split1} \in \{90\mathrm{split1}, 150\mathrm{split1} \}$, $\mathds{1}$ is a vector of ones, and $\Delta \ell = 20$. This calculation is performed at every $\ell$ within windows that overlap, and then binned as described previously (note that the binning in the harmonic ILC computation at each $\ell$ is distinct from the binning of the final HILC spectra for use in the data vector). We then compute weights for the HILC Compton-$y$ map using the second noise split: 
\begin{equation}
    w^{i,\hat{y}\mathrm{split2}}_\ell = \frac{ \left( \hat{R}_{\ell}^{-1} \right)_{ij} g^j_y }{ \left( \hat{R}_{\ell}^{-1} \right)_{km} g^k_y g^m_y}
    \qquad \text{with} \qquad 
    \left(\hat{R}_{\ell} \right)_{ij}= \sum_{\ell' = \ell-\Delta \ell /2}^{\ell+\Delta \ell /2} \frac{2\ell'+1}{4\pi} C_{\ell'}^{i\mathrm{split2},j\mathrm{split2}} \, .
\end{equation}
Then,
\begin{equation}
    C_\ell^{\hat{T} \mathrm{ split1}, \hat{y} \mathrm{split2}} = \sum_{i,j} w^{i,\hat{T}\mathrm{split1}}_\ell w^{j,\hat{y}\mathrm{split2}}_\ell C_\ell^{i\mathrm{split1},j\mathrm{split2}} \, .
\end{equation}
After computing the spectrum independently for each multipole, $\ell$-space binning is performed, giving $C_b^{\hat{T} \mathrm{ split1}, \hat{y} \mathrm{split2}}$. The computation of $C_b^{\hat{T} \mathrm{ split1}, \hat{T} \mathrm{split2}}$, $C_b^{\hat{y} \mathrm{ split1}, \hat{T} \mathrm{split2}}$, and $C_b^{\hat{y} \mathrm{ split1}, \hat{y}\mathrm{split2}}$ are done analogously. In Appendix~\ref{app:maps} we verify that this analytic computation of HILC spectra yields the same results as producing HILC maps and then computing the spectra from the maps.

We use 2000 simulations for the HILC Gaussian likelihood pipelines (described in Sec.~\ref{sec.method_hilc_gaussian_lkl}). We have two options for  computing the weights in the pipeline: 
\begin{enumerate}
    \item Fixed weights: Compute the HILC weights once from some fiducial template multifrequency power spectra. Apply these same weights to every simulation. For the remainder of the paper, we refer to this as the ``fixed weights" or ``weights once" HILC variant.
    \item Varying weights: Compute the HILC weights separately for every simulation, using the specific multifrequency power spectra from that simulation. For the remainder of the paper, we refer to this as the ``varying weights" or ``weights vary" HILC variant.
\end{enumerate}
When the weights vary on a per-simulation basis, we ensure that the ILC bias is mitigated by omitting the central $\ell$ value in each bin when computing the weights, i.e., by setting
\begin{equation}
    \left(\hat{R}_{\ell} \right)_{ij}= \left( \sum_{\ell' = \ell-\Delta \ell /2}^{\ell+\Delta \ell /2} \frac{2\ell'+1}{4\pi} C_{\ell'}^{ij} \right) - \frac{2\ell+1}{4\pi} C_{\ell}^{ij}\, .
\end{equation}
This heavily reduces chance correlations between the signal and noise/contaminants, which thereby reduces the ILC bias (this is analogous to a similar trick developed for NILC in Ref.~\cite{Coulton2023ACT}).  The two variants of the HILC pipeline have different implications for parameter dependence in the Gaussian likelihood. This is discussed further below in Sec.~\ref{sec.method_hilc_gaussian_lkl}.

\subsection{Gaussian Likelihood}
\label{sec.method_hilc_gaussian_lkl}
The Gaussian likelihood is given by 

\begin{align}
    \label{eq.hilc_lkl}
    \ln \mathcal{L}(A_{\rm CMB}, A_{\rm ftSZ}) &= -\frac{1}{2} \left( C_{b_1}^{\hat{p}\hat{q}}(A_{\rm CMB}, A_{\rm ftSZ}) - C_{b_1}^{\hat{p}\hat{q}, \mathrm{data}} \right)  \left(\mathrm{Cov}_{\hat{p}\hat{q} b_1, \hat{r}\hat{s} b_2}\right)^{-1} \left( C_{b_2}^{\hat{r}\hat{s}}(A_{\rm CMB}, A_{\rm ftSZ}) - C_{b_2}^{\hat{r}\hat{s}, \mathrm{data}} \right) \, ,
\end{align}
where summation is implied over repeated indices $\hat{p}$, $\hat{q}$, $\hat{r}$, $\hat{s}$, $b_1$, and $b_2$ on the RHS. Here $C_{b_1}^{\hat{p}\hat{q}}, C_{b_2}^{\hat{r}\hat{s}} \in \{ C_b^{\hat{T} \mathrm{ split1}, \hat{T} \mathrm{split2}}, C_b^{\hat{T} \mathrm{ split1}, \hat{y} \mathrm{split2}}, C_b^{\hat{y} \mathrm{ split1}, \hat{T} \mathrm{split2}},  C_b^{\hat{y} \mathrm{ split1}, \hat{y} \mathrm{split2}} \}$ and $\mathrm{Cov}_{\hat{p}\hat{q} b_1,\hat{r}\hat{s} b_2} \equiv \mathrm{Cov}(C_{b_1}^{\hat{p}\hat{q}}, C_{b_2}^{\hat{r}\hat{s}})$ is the covariance matrix of the HILC power spectra, computed directly from 2000 independent simulations generated at the fiducial parameter values (i.e., there is no assumption of a Gaussian covariance matrix). A correction factor is applied to the inverse of the HILC power spectrum covariance matrix, as is done in Sec.~\ref{sec.method_multifrequency_power_spec_gaussian_lkl} for the multifrequency power spectrum covariance matrix \cite{Hartlap:2006}. Note also that we have included off-diagonal $\ell$-bins in the covariance matrix and likelihood.

\subsubsection{Analytic Parameter Dependence}
\label{sec.method_hilc_gaussian_lkl_analytic}
When using fixed weights in the HILC pipeline, the weights do not depend on the particular fluctuations of any single simulated sky realization. Thus, we can analytically write down the parameter dependence as (ignoring binning for now)
\begin{equation}
    \label{eq.hilc_data_vec_no_binning}
    C_{\ell}^{\hat{p}\hat{q}} = \sum_{i,j} w^{i,\hat{p}}_{\ell} w^{j,\hat{q}}_{\ell} C_{\ell}^{ij} (A_{\mathrm{CMB}}, A_{\mathrm{ftSZ}}) = \sum_{i,j} w^{i,\hat{p}}_{\ell} w^{j,\hat{q}}_{\ell} (A_{\rm CMB} C_{\ell}^{\mathrm{CMB}} + A_{\rm ftSZ} g^i_y g^j_y C_{\ell}^{\mathrm{ftSZ}}) \, ,
\end{equation}
where we have used Eq.~\eqref{eq.multifrequency_data_vec} in the second equality. Including binning, we obtain 
\begin{equation}
    \label{eq.hilc_data_vec_binned}
    C_{b_1}^{\hat{p}\hat{q}} = \frac{1}{N_{b_1}} \sum_{\ell \in b_1} \sum_{i,j} w^{i,\hat{p}}_{\ell} w^{j,\hat{q}}_{\ell} C_{\ell}^{ij} (A_{\mathrm{CMB}}, A_{\mathrm{ftSZ}}) = \frac{1}{N_{b_1}} \sum_{\ell \in b_1} \sum_{i,j} w^{i,\hat{p}}_{\ell} w^{j,\hat{q}}_{\ell} (A_{\rm CMB} C_{\ell}^{\mathrm{CMB}} + A_{\rm ftSZ} g^i_y g^j_y C_{\ell}^{\mathrm{ftSZ}}) \, ,
\end{equation}
where $N_{b_1}$ is the number of multipole values $\ell$ contained in bin $b_1$ (it is \emph{not} the total number of modes ($\ell,m$)). 

However, this simple parameter dependence is not correct in the varying-weights case. In that case, the weights respond to the particular fluctuations in the sky signal realized in each simulation.  Thus, they are functions of the (randomly fluctuating) parameter amplitude values of individual simulations. Determining the parameter dependence in the varying-weights case is the subject of Sec.~\ref{sec.method_hilc_gaussian_lkl_sr} below.

\subsubsection{Parameter Dependence from Symbolic Regression}
\label{sec.method_hilc_gaussian_lkl_sr}
Our goal in this section is to write 
\begin{equation}
    \label{eq.sr_f_def}
    C^{\hat{p}\hat{q}}_{b}(A_{\mathrm{CMB}}, A_{\mathrm{ftSZ}}) = f^{\hat{p}\hat{q}}_{b}(A_{\mathrm{CMB}}, A_{\mathrm{ftSZ}}) C^{\hat{p}\hat{q}}_{b}(A_{\mathrm{CMB}} = 1 = A_{\mathrm{ftSZ}})
\end{equation}
and determine a function $f^{\hat{p}\hat{q}}_{b}(A_{\mathrm{CMB}}, A_{\mathrm{ftSZ}})$ for each $\hat{p}\hat{q} \in \{ \hat{T}\mathrm{split1}\hat{T}\mathrm{split2}, \hat{T}\mathrm{split1}\hat{y}\mathrm{split2}, \hat{y}\mathrm{split1}\hat{T}\mathrm{split2}, \hat{y}\mathrm{split1}\hat{y}\mathrm{split2} \}$ and bin $b$. The superscripts and subscripts on $f$ only denote that it is a quantity to be determined separately for each $\hat{p}\hat{q}$ and $b$; however, $f$ is just some function, not a power spectrum. Moreover, there is no implied summation over $\hat{p}$, $\hat{q}$, or $b$ in the equation. To avoid assuming a specified form of $f$, we find these functions with symbolic regression, using \verb|PySR| \cite{pysr}. For our regressor, we allow the binary operations of addition, subtraction, multiplication, and division; unary operations of exponentiation, squaring, and cubing; and a custom inverse operation. We use an L2-loss function that does not in itself penalize complexity since we simply need smooth functions that describe the simulations well, rather than concise equations representing any fundamental law of physics. However, we do impose a complexity limit to avoid overfitting by setting the maximum size of our expression to 12. 

For each $\hat{p}\hat{q}$ and bin $b$, the values that we feed into the regressor come from scaling various combinations of components. We define four ``scaling factors" around the fiducial parameter value of 1: 0.9, 0.99, 1.01, and 1.1. First, we find the average HILC spectra over 50 simulations and use these to compute $$C_b^{\hat{p}\hat{q}}(A_{\mathrm{CMB}}=1=A_{\mathrm{ftSZ}}).$$ As an example, consider the scaling factor 0.9. Using the same map-level realizations that were used to determine $C_b^{\hat{p}\hat{q}}(A_{\mathrm{CMB}}= 1 = A_{\mathrm{ftSZ}})$, we multiply the CMB and/or the ftSZ map by 0.9 for each of the 50 simulations. The ILC weights (in the varying weights case) also change as a result. Then we compute
$$C_b^{\hat{p}\hat{q}}(A_{\mathrm{CMB}}=0.81, A_{\mathrm{ftSZ}}=1), \; C_b^{\hat{p}\hat{q}}(A_{\mathrm{CMB}}=1, A_{\mathrm{ftSZ}}=0.81), \; \text{and } C_b^{\hat{p}\hat{q}}(A_{\mathrm{CMB}}=0.81, A_{\mathrm{ftSZ}}=0.81) \, .$$
Note that the scaled parameters are set to $0.81$ since multiplying by $0.9$ at the map level corresponds to multiplying by $0.9^2=0.81$ at the power spectrum level. For this example, the data points $(\vec{x},y)$ that go into the regressor are then 
\begin{align}
    \vec{x}=(A_{\mathrm{CMB}},A_{\mathrm{ftSZ}})=(0.81,1) &, \qquad y=C_b^{\hat{p}\hat{q}}(A_{\mathrm{CMB}}=0.81,A_{\mathrm{ftSZ}}=1)/ C_b^{\hat{p}\hat{q}}(A_{\mathrm{CMB}}=1=A_{\mathrm{ftSZ}}) \, , \nonumber \\
    \vec{x}=(A_{\mathrm{CMB}},A_{\mathrm{ftSZ}})=(1,0.81) &, \qquad y=C_b^{\hat{p}\hat{q}}(A_{\mathrm{CMB}}=1,A_{\mathrm{ftSZ}}=0.81)/ C_b^{\hat{p}\hat{q}}(A_{\mathrm{CMB}}=1=A_{\mathrm{ftSZ}}) \, , \text{ and} \nonumber \\
    \vec{x}=(A_{\mathrm{CMB}},A_{\mathrm{ftSZ}})=(0.81,0.81) &, \qquad y=C_b^{\hat{p}\hat{q}}(A_{\mathrm{CMB}}=0.81=A_{\mathrm{ftSZ}})/ C_b^{\hat{p}\hat{q}}(A_{\mathrm{CMB}}=1=A_{\mathrm{ftSZ}}) \, . \nonumber
\end{align}
We repeat the same procedure with the other scaling factors, feeding the $(\vec{x},y)$ pairs from all scaling factors into the regressor together. Then by definition of $\vec{x}$ and $y$, the regressor learns $f^{\hat{p}\hat{q}}_b$.

This procedure is used to determine the parameter dependence for the HILC varying-weights Gaussian-likelihood pipeline. However, it can also be used on the HILC fixed-weights Gaussian-likelihood pipeline. Since we know the analytic parameter dependence in the fixed-weights case, for that scenario we can compare the posteriors obtained using analytic parameter dependence to the posteriors obtained using SR as a cross-check of the validity of the SR approach. This serves as an important validation step for the varying-weights HILC Gaussian-likelihood pipeline, as well as the NILC Gaussian-likelihood pipeline that is described in Sec.~\ref{sec.method_nilc_gaussian_lkl}. Sample equations and parameter dependence learned via SR are shown in Appendix~\ref{app:sr}. Note that, because of the way we define the parameter dependence function $f^{\hat{p}\hat{q}}_b$ in Eq.~\eqref{eq.sr_f_def}, we cannot directly compare the equations obtained from symbolic regression to the analytic parameter dependence in Eq.~\eqref{eq.hilc_data_vec_binned}. We could have instead defined the symbolic regression procedure such that it could recover that same analytic parameter dependence, but the computational cost would be greater since we would have to perform symbolic regression multiple times for each $\hat{p}$, $\hat{q}$, and $b$. 

\subsubsection{MLE, Fisher Matrix, and MCMC}
\label{sec.method_hilc_gaussian_lkl_posteriors}

As for the multifrequency power spectrum Gaussian likelihood, we obtain parameter posteriors from HILC map power spectra via MLE, the Fisher matrix, and an MCMC algorithm. For MLE, we numerically find the maximum-likelihood point of the likelihood function in Eq.~\eqref{eq.hilc_lkl} for each of the 2000 simulations. Specifically, for each simulation, we compute the HILC spectra from that simulation to use as $C_{b_1}^{\hat{p}\hat{q},\mathrm{data}}$ and $C_{b_2}^{\hat{r}\hat{s},\mathrm{data}}$. We then find the maximum-likelihood $(A_{\mathrm{CMB}}, A_{\mathrm{ftSZ}})$ point for every simulation using Nelder-Mead minimization \cite{Nelder-Mead} of the negative log likelihood. We run the minimization routine using three starting points: $(A_{\mathrm{CMB}}, A_{\mathrm{ftSZ}}) = (1.0, 1.0)$, $(0.5, 0.5)$, and $(1.5, 1.5)$. We then use the global minimum from these results as our maximum-likelihood point $(\hat{A}_{\mathrm{CMB}}, \hat{A}_{\mathrm{ftSZ}})$.

The Fisher matrix in this approach is defined as
\be
\label{eq.Fisher_hilc}
    F_{\hat{A}_{\alpha} \hat{A}_{\beta}} = 
    -\left. \left( \frac{ \partial^2 \ln{\mathcal{L}} }{ \partial A_{\alpha} A_{\beta} } \right) \right|_{\hat{A}_{\alpha} \hat{A}_{\beta}}
    =  \left. \frac{\partial C_{b_1}^{\hat{p}\hat{q}}}{\partial A_{\alpha}} (\mathrm{Cov}^{-1}_{\hat{p}\hat{q}b_1,\hat{r}\hat{s}b_2}) \frac{\partial C_{b_2}^{\hat{r}\hat{s}}}{\partial A_{\beta}}   \right|_{A_{\alpha}=A_{\beta}=1} \, ,
\ee
where there is implied summation over $\hat{p}$, $\hat{q}$, $\hat{r}$, $\hat{s}$, $b_1$, and $b_2$. For the case of fixed weights with analytic parameter dependence, we compute the derivatives in Eq.~\eqref{eq.Fisher_hilc} analytically (cf.~Eq.~\eqref{eq.hilc_data_vec_binned}). For varying weights and fixed weights with parameter dependence determined via SR, we compute the derivatives numerically with finite differences.

Finally, for the MCMC algorithm we use \verb|emcee| \cite{emcee} with the likelihood in Eq.~\eqref{eq.hilc_lkl} and a mock observation consisting of the mean of simulated HILC auto- and cross-spectra over all simulations. The remaining set-up of the MCMC algorithm is the same as described in Sec.~\ref{sec.method_multifrequency_power_spec_gaussian_lkl}.

\subsection{Likelihood-Free Inference}
\label{sec.method_hilc_lfi}
We follow a similar procedure as for the LFI in the multifrequency power spectrum case, described in Sec.~\ref{sec.method_multifrequency_power_spec_lfi}.
For the prior, we set a uniform prior centered on 1 with lower bound $(1-5\sigma_{\hat{A}_{\mathrm{CMB}}}, 1-5\sigma_{\hat{A}_{\mathrm{ftSZ}}})$ and upper bound $(1+5\sigma_{\hat{A}_{\mathrm{CMB}}}, 1+5\sigma_{\hat{A}_{\mathrm{ftSZ}}})$, where $\sigma_{\hat{A}_{\mathrm{CMB}}}$ and $\sigma_{\hat{A}_{\mathrm{ftSZ}}}$ are the 1D marginalized errors obtained from the multifrequency Gaussian likelihood on $\hat{A}_{\mathrm{CMB}}$ and $\hat{A}_{\mathrm{tSZ}}$, respectively. Note that the prior is kept the same as the prior in the multifrequency LFI for fair comparison of the resulting posteriors.

As for the HILC Gaussian likelihood, we perform LFI with two versions of the HILC method: one in which the weights are computed once from some fiducial template generated at $A_{\mathrm{CMB}}=1=A_{\mathrm{ftSZ}}$, and one in which the weights are computed independently for each output of the simulator. For the simulator, we draw $A_{\mathrm{CMB}}$ and $A_{\mathrm{ftSZ}}$ from the prior distribution. Then for each simulation, we multiply the input CMB map by a factor of $\sqrt{A_{\mathrm{CMB}}}$ and the ftSZ map by a factor of $\sqrt{A_{\mathrm{ftSZ}}}$. We then proceed as usual with creating maps of two noise splits for each frequency and performing the HILC operations to construct the data vector comprising the concatenation of $C_b^{\hat{T} \mathrm{ split1}, \hat{T} \mathrm{split2}}$, $C_b^{\hat{T}\mathrm{ split1}, \hat{y} \mathrm{split2}}$, $C_b^{\hat{y} \mathrm{ split1}, \hat{T} \mathrm{split2}}$, and $C_b^{\hat{y} \mathrm{ split1}, \hat{y} \mathrm{split2}}$. Thus, the data vector has a total length of 40 since there are 10 bins in each of the concatenated spectra. The remainder of the LFI setup is the same as described in Sec.~\ref{sec.method_multifrequency_power_spec_lfi}.

\section{Needlet ILC Component Map Power Spectra as Summary Statistics}
\label{sec.method_nilc}

\subsection{Gaussian Likelihood}
\label{sec.method_nilc_gaussian_lkl}

We follow a procedure similar to that described in Sec.~\ref{sec.method_hilc_gaussian_lkl}. Our data vector for the NILC component map auto- and cross-power spectra consists of $C_b^{\hat{T} \mathrm{ split1}, \hat{T} \mathrm{split2}}$, $C_b^{\hat{T} \mathrm{ split1}, \hat{y} \mathrm{split2}}$, $C_b^{\hat{y} \mathrm{ split1}, \hat{T} \mathrm{split2}}$, and $C_b^{\hat{y} \mathrm{ split1}, \hat{y}\mathrm{split2}}$, where $\hat{T}\mathrm{split1}$ signifies a NILC CMB map produced using the frequency maps with the first noise split, $\hat{y}\mathrm{split1}$ signifies a NILC Compton-$y$ map produced using the frequency maps with the first noise split, and similarly for the second split. The subscript $b$ denotes the multipole bin, where we use 10 bins from $\ell=2$ to $\ell=250$ (one bin contains 24 multipoles, and all other bins contain 25 multipoles). 

In the HILC case, we are able to easily compute the power spectra analytically. However, this is not the case for NILC, as shown in Ref.~\cite{surrao2024_nilc1}, due to the non-trivial contributions of of three- and four-point functions of the components and weight maps. Thus, we build NILC maps from simulations and compute the power spectra directly from these. To produce the NILC maps and weight maps, we use \verb|pyilc|\footnote{\url{https://github.com/jcolinhill/pyilc}} \cite{McCarthy:2023cwg, McCarthy:2023hpa}. We use needlet filters that are the difference of successive Gaussians, where the Gaussians have full width at half maximum (FWHM) values of 300, 120, and 60 arcmin, resulting in four needlet filter scales. The fourth filter is modified slightly such that the filters satsify the needlet filter power preservation constraint, i.e. that $\sum_{(n)} \left( h_\ell^{(n)} \right) ^2 = 1$ at each $\ell$. The filters are shown in Fig.~\ref{fig:needlet_filters}. Examples of the resulting NILC component maps and weight maps are shown in Appendix \ref{app:maps}.

\begin{figure}[t]
    \centering
    \includegraphics[width=0.5\textwidth]{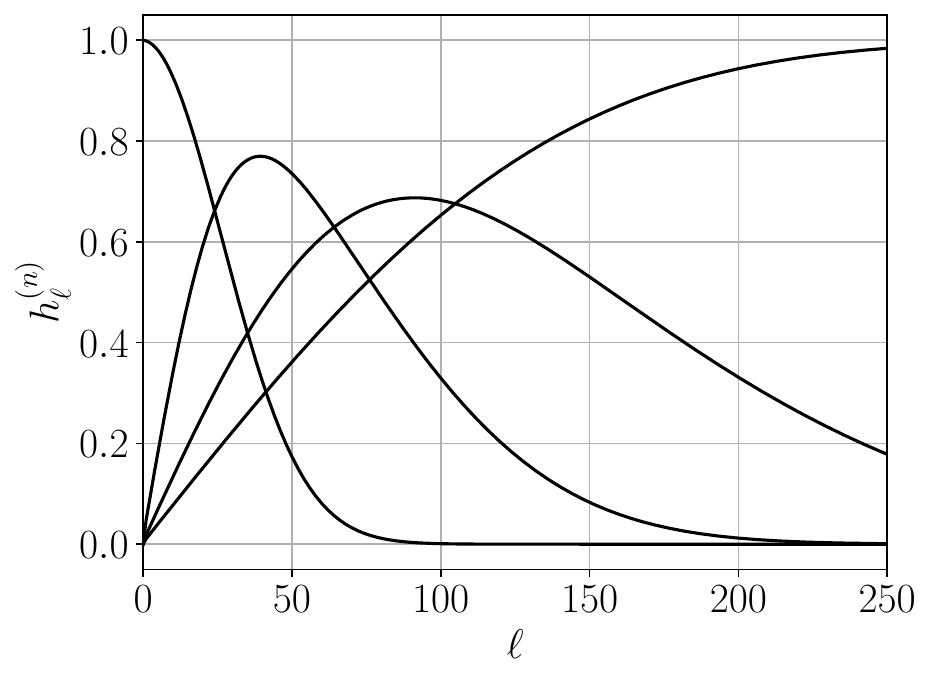}
    \caption{Needlet filters $h_\ell^{(n)}$ used in our simulations for the NILC pipeline, found by taking the differences of successive Gaussians of FWHM 300, 120, and 60 arcmin.}
    \label{fig:needlet_filters}
\end{figure}

In the HILC case, we implement two variants: one with fixed ILC weights (i.e., weights that are not re-computed for every simulation realization) and one with varying ILC weights (i.e., weights that are re-computed for every realization). In the NILC case, we only use varying weights, meaning that the weight maps are determined separately for every simulation. This is necessary so that the NILC algorithm can capture meaningful spatially varying information within each simulation. In particular, on a single simulation, the pipeline is as follows: 
\begin{enumerate}
    \item Simulate four frequency maps (maps of two noise splits for each of the two frequencies).
    \item For each split, run \verb|pyilc| to build a NILC CMB-preserved map (e.g., to build a CMB NILC map for the first split, feed in the 90 GHz Split 1 and 150 GHz Split 1 maps as input to the code and specify preservation of the CMB component). Repeat the same procedure for the (amplified) Compton-$y$ NILC map. Thus, for each simulation we have four NILC maps: $\hat{T}\mathrm{split1}$, $\hat{T}\mathrm{split2}$, $\hat{y}\mathrm{split1}$, and $\hat{y}\mathrm{split2}$.
    \item Compute the elements of the data vector (i.e., the power spectra): $C_b^{\hat{T} \mathrm{ split1}, \hat{T} \mathrm{split2}}$, $C_b^{\hat{T} \mathrm{ split1}, \hat{y} \mathrm{split2}}$, $C_b^{\hat{y} \mathrm{ split1}, \hat{T} \mathrm{split2}}$, and $C_b^{\hat{y} \mathrm{ split1}, \hat{y}\mathrm{split2}}$.
\end{enumerate}

We then use the same Gaussian likelihood as in Eq.~\eqref{eq.hilc_lkl}, except that the spectra are now NILC spectra instead of HILC spectra. To determine the parameter dependence $C_{b}^{\hat{p}\hat{q}}(A_{\mathrm{CMB}}, A_{\mathrm{ftSZ}})$, we follow the same symbolic regression procedure that is described in Sec.~\ref{sec.method_hilc_gaussian_lkl_sr}. In Appendix \ref{app:sr}, we show examples of the expressions and parameter dependence we obtain from symbolic regression, and compare them to those obtained in the HILC case.

\begin{figure}[t]
    \centering
    \includegraphics[width=0.95\textwidth]{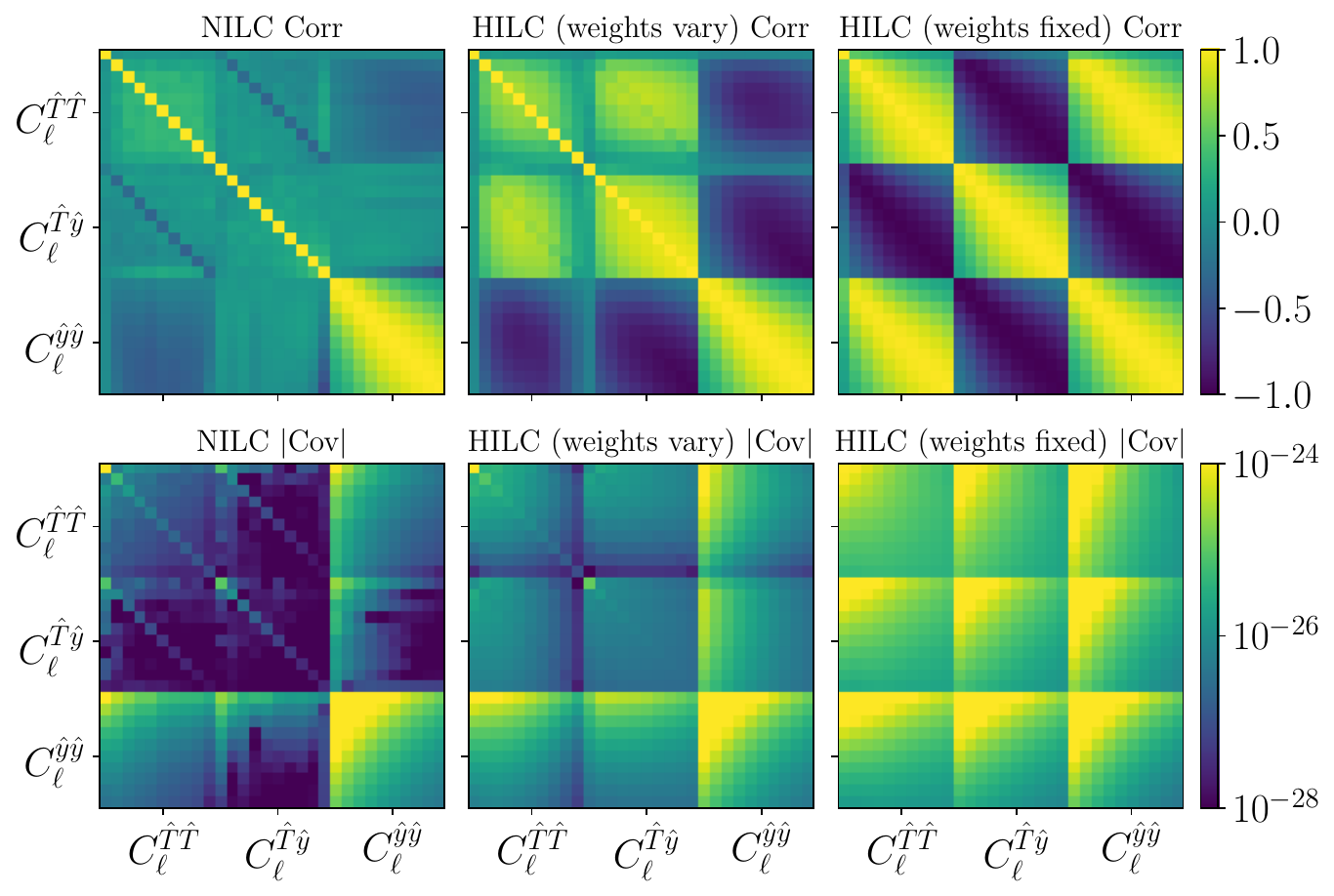}
    \caption{Correlation matrices (top row) and absolute value of covariance matrices (bottom row) computed via 2000 simulations, as used in our Gaussian likelihoods. These matrices are shown for the NILC pipeline (left), HILC with varying weights pipeline (middle), and HILC with fixed weights pipeline (right). Within each subplot, there is a grid of 9 squares, corresponding to elements of the covariance and correlation among $C_\ell^{\hat{T}\hat{T}}$, $C_\ell^{\hat{T}\hat{y}}$, and $C_\ell^{\hat{y}\hat{y}}$ for each of the 10 multipole bins increasing from left to right and top to bottom. Here $C_\ell^{\hat{T}\hat{y}}$ corresponds to $C_\ell^{\hat{T}\mathrm{split1},\hat{y}\mathrm{split2}}$ and we omit the other split combination for concision since it behaves similarly. The $\hat{T}$ maps are in units of $\mathrm{K}_{\mathrm{CMB}}$, while the $\hat{y}$ maps are in dimensionless Compton-$y$ units.}
    \label{fig:cov}
\end{figure}

\begin{figure}[htb]
    \centering
    \includegraphics[width=0.75\textwidth]{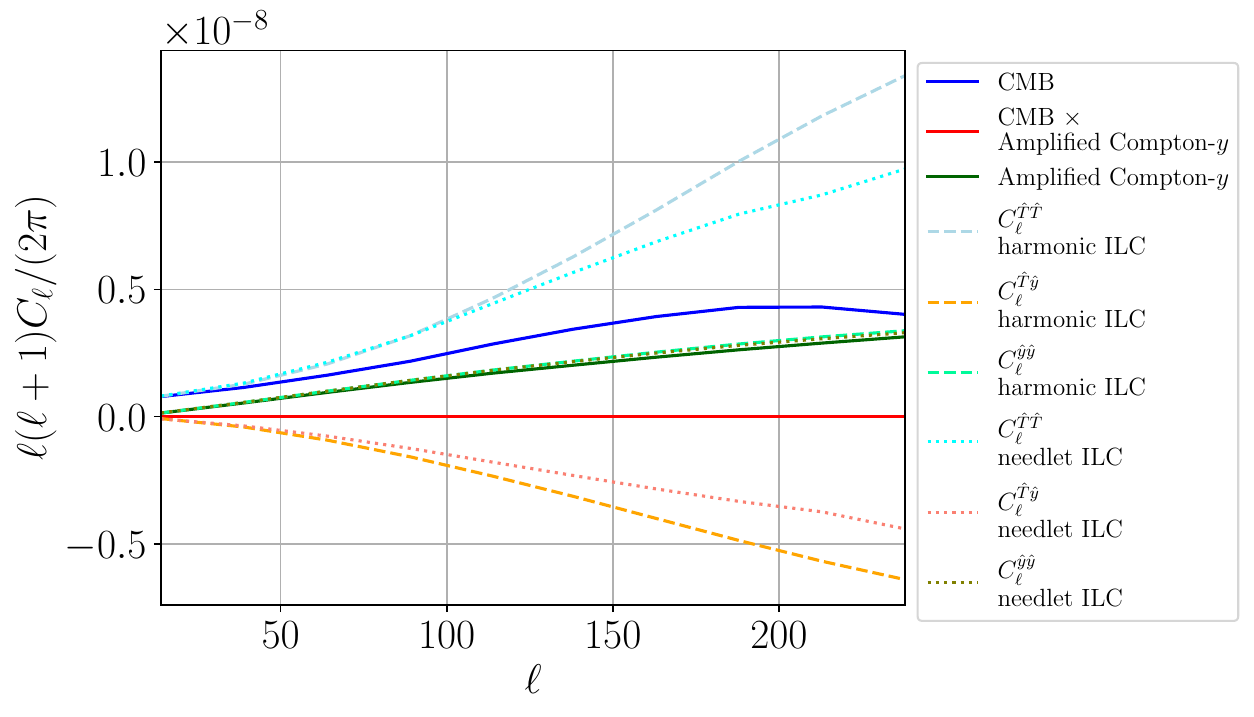}
    \caption{Mean ILC spectra over 2000 simulations, for NILC and HILC (with varying weights, i.e., weights determined separately for each of the simulations). Solid curves show the ground truth spectra for the CMB (blue), CMB $\mathrm \times$ amplified Compton-$y$ (red), and amplified Compton-$y$ (green). Dashed curves show HILC spectra: $C_\ell^{\hat{T}\hat{T}}$ (light blue), $C_\ell^{\hat{T}\hat{y}}$ (orange), and $C_\ell^{\hat{y}\hat{y}}$ (lime green). Dotted curves show NILC spectra: $C_\ell^{\hat{T}\hat{T}}$ (cyan), $C_\ell^{\hat{T}\hat{y}}$ (salmon), and $C_\ell^{\hat{y}\hat{y}}$ (olive). Here $\hat{T}$ represents a CMB-preserved ILC map, and $\hat{y}$ represents an ILC map that preserves the tSZ signal. Each of the spectra are computed using noise splits (here $C_\ell^{\hat{T}\hat{y}}$ is really $C_\ell^{\hat{T}\mathrm{split1},\hat{y}\mathrm{split2}}$ and the other split combination is omitted for concision since it behaves similarly). The $\hat{T}$ maps are in units of $\mathrm{K}_{\mathrm{CMB}}$, and the $\hat{y}$ maps are in dimensionless Compton-$y$ units. }
    \label{fig:ILC_spectra}
\end{figure}

We compute the covariance matrix used in the Gaussian likelihood directly via 2000 simulations of the above pipeline. A correction factor is applied to the inverse of the NILC power spectrum covariance matrix, as is done in Sec.~\ref{sec.method_multifrequency_power_spec_gaussian_lkl} for the multifrequency power spectrum covariance matrix \cite{Hartlap:2006}. We then proceed with the MLE, Fisher matrix, and MCMC procedures, as described in Sec.~\ref{sec.method_hilc_gaussian_lkl_posteriors}. In Fig.~\ref{fig:cov}, we show the correlation matrix and covariance matrix from these NILC simulations and compare them to the HILC correlation and covariance matrices for both the fixed-weights and varying-weights cases (described in Sec.~\ref{sec.method_hilc}). In all cases, the non-Gaussianity of the tSZ field is evident from the off-diagonal contributions to $\mathrm{Cov}(C_\ell^{\hat{y}\hat{y}}, C_\ell^{\hat{y}\hat{y}})$, and the propagation of that field as a contaminant to the $\mathrm{Cov}(C_\ell^{\hat{T}\hat{T}}, C_\ell^{\hat{T}\hat{T}})$ covariance can be seen. However, we see that this contamination is lowest in the NILC case, where the $\mathrm{Cov}(C_\ell^{\hat{T}\hat{T}}, C_\ell^{\hat{T}\hat{T}})$ is roughly diagonal. We also note that the varying-weights HILC does better than the fixed-weights HILC, as expected since the varying weights can account for (isotropic) harmonic-space fluctuations in individual simulations. As further confirmation that NILC better cleans contaminants, we show the mean HILC (in the varying-weights case) and mean NILC spectra from 2000 simulations in Fig.~\ref{fig:ILC_spectra}. From this figure, it is evident that NILC better suppresses contamination and recovers spectra that are closer to the underlying ground truth. 

\subsection{Likelihood-Free Inference}
\label{sec.method_nilc_lfi}
We follow a similar procedure to that used for the LFI in the multifrequency-power-spectra and HILC-power-spectra cases, described in Sec.~\ref{sec.method_multifrequency_power_spec_lfi} and \ref{sec.method_hilc_lfi}, respectively. For the prior, we set a uniform prior centered on 1 with lower bound $(1-5\sigma_{\hat{A}_{\mathrm{CMB}}}, 1-5\sigma_{\hat{A}_{\mathrm{ftSZ}}})$ and upper bound $(1+5\sigma_{\hat{A}_{\mathrm{CMB}}}, 1+5\sigma_{\hat{A}_{\mathrm{ftSZ}}})$, where $\sigma_{\hat{A}_{\mathrm{CMB}}}$ and $\sigma_{\hat{A}_{\mathrm{ftSZ}}}$ are the 1D marginalized errors obtained from the multifrequency Gaussian likelihood on $\hat{A}_{\mathrm{CMB}}$ and $\hat{A}_{\mathrm{tSZ}}$, respectively. Note that the prior is kept the same as the prior in the multifrequency and HILC LFI for fair comparison of the resulting posteriors.

For the simulator, we draw $A_{\mathrm{CMB}}$ and $A_{\mathrm{ftSZ}}$ from the prior distribution. Then for each simulation, we multiply the input CMB map by a factor of $\sqrt{A_{\mathrm{CMB}}}$ and the ftSZ map by a factor of $\sqrt{A_{\mathrm{ftSZ}}}$. We then proceed as in the Gaussian-likelihood case with creating maps of two noise splits for each frequency and performing the NILC operations to construct the data vector comprising the concatenation of $C_b^{\hat{T} \mathrm{ split1}, \hat{T} \mathrm{split2}}$, $C_b^{\hat{T}\mathrm{ split1}, \hat{y} \mathrm{split2}}$, $C_b^{\hat{y} \mathrm{ split1}, \hat{T} \mathrm{split2}}$, and $C_b^{\hat{y} \mathrm{ split1}, \hat{y} \mathrm{split2}}$. Thus, the data vector has a total length of 40 since there are 10 bins in each of the concatenated spectra. The remainder of the LFI setup is the same as described in Sec.~\ref{sec.method_multifrequency_power_spec_lfi}.

\section{When are the harmonic ILC and multifrequency power spectrum inference methods analytically equivalent?}
\label{sec.method_hilc_gaussian_lkl_proof}

\subsection{General Set-Up}

In this section, we perform analytic calculations to compare parameter covariances derived from HILC map power spectra and template fitting to multifrequency auto- and cross-power spectra.  We determine the conditions under which these two inference methods are exactly equivalent, and explicitly prove this result.  For all calculations here, we assume a Gaussian likelihood for both approaches.  We also assume no $\ell$-space binning, but the results also hold in the limit of small bin widths. Finally, for clarity and concision we do not use different noise splits of the data, though the results hold in that case as well.

We prove the equivalence of the parameter Fisher matrices, and thus the final parameter covariance matrices, at a single multipole $\ell$ (or for a single narrow bandpower). This immediately generalizes to fitting all multipoles simultaneously, as long as the power spectrum covariance matrix is diagonal in $\ell$. The Fisher matrix for the multifrequency approach is given by Eq.~\eqref{eq.Fisher_mult}:
\be
    F^{\mathrm{mult}}_{\hat{A}_{\alpha} \hat{A}_{\beta}} = 
 \sum_{ij} \sum_{km} \left. \frac{\partial C_{\ell}^{ij}}{\partial A_{\alpha}} \left(\mathrm{Cov}^{-1}\right)^{ij,km}_\ell \frac{\partial C_{\ell}^{km}}{\partial A_{\beta}}   \right|_{A_{\alpha}=1=A_{\beta}} \, ,
\ee
where $ij,km$ can each take on $N_{\mathrm{freq}}(N_{\mathrm{freq}}+1)/2$ values. For example, if there are two frequency channels, then $ij,km$ can each take on three values: one for each frequency-frequency auto-spectrum, and one for the frequency-frequency cross-spectrum.  Since we consider only a single $\ell$ here, there is no sum over $\ell,\ell^\prime$, as would be present in the general form of the above equation.  The Fisher matrix for the HILC approach is given by Eq.~\eqref{eq.Fisher_hilc}:
\be
    F^{\mathrm{HILC}}_{\hat{A}_{\alpha} \hat{A}_{\beta}} 
    = \sum_{\hat{p}\hat{q}} \sum_{\hat{r}\hat{s}} \left. \frac{\partial C_{\ell}^{\hat{p}\hat{q}}}{\partial A_{\alpha}} \left(\mathrm{Cov}^{-1}\right)^{\hat{p}\hat{q},\hat{r}\hat{s}}_\ell \frac{\partial C_{\ell}^{\hat{r}\hat{s}}}{\partial A_{\beta}}   \right|_{A_{\alpha}=1=A_{\beta}} \, ,
\ee
where the number of values that $\hat{p}\hat{q}, \hat{r}\hat{s}$ can take on is more complicated than in the multifrequency case and is examined in detail in the next subsection. The power spectra of HILC maps are easy to model. In particular, from Eq.~\eqref{eq.hilc_data_vec_no_binning}, we have 
\begin{equation}
    \label{eq.HILCderiv}
     \frac{\partial C_{\ell}^{\hat{p}\hat{q}}}{\partial A_{\alpha}}  = \sum_{i,j} w^{i,\hat{p}}_{\ell} w^{j,\hat{q}}_{\ell}  \frac{\partial C_{\ell}^{ij}}{\partial A_{\alpha}} \qquad \text{and} \qquad
     \frac{\partial C_{\ell}^{\hat{r}\hat{s}}}{\partial A_{\beta}}  = \sum_{k,l} w^{k,\hat{r}}_{\ell} w^{l,\hat{s}}_{\ell_2}  \frac{\partial C_{\ell}^{kl}}{\partial A_{\beta}}   \,.
\end{equation}
The covariance matrix for the HILC component map auto- and cross-power spectra is (at a single $\ell$):
\begin{equation}
    \label{eq.HILCcov}
    \mathrm{Cov}^{\hat{p}\hat{q}, \hat{r}\hat{s}}_\ell = \mathrm{Cov}(C_{\ell}^{\hat{p}\hat{q}}, C_{\ell}^{\hat{r}\hat{s}}) = \sum_{i,j,k,m} w^{i,\hat{p}}_{\ell} w^{j,\hat{q}}_{\ell} w^{k,\hat{r}}_{\ell} w^{m,\hat{s}}_{\ell} \mathrm{Cov} (C_{\ell}^{ij}, C_{\ell}^{km}) \, ,
\end{equation}
where here we are summing over individual frequencies, i.e., $i,j,k,l$ each take on $N_{\mathrm{freq}}$ possible values.
Then we have 
\begin{align}
    \label{eq.HILC_Fisher_analytic}
    F_{\hat{A}_\alpha \hat{A}_\beta}^{\mathrm{HILC}} &= \left. \sum_{\hat{p}\hat{q},\hat{r}\hat{s}} \left(\sum_{i,j} w^{i,\hat{p}}_{\ell} w^{j,\hat{q}}_{\ell}  \frac{\partial C_{\ell}^{ij}}{\partial A_{\alpha}}\right) 
    \left( \sum_{i,j,k,l} w^{i,\hat{p}}_{\ell} w^{j,\hat{q}}_{\ell} w^{k,\hat{r}}_{\ell} w^{l,\hat{s}}_{\ell} \mathrm{Cov} (C_{\ell}^{ij}, C_{\ell}^{kl}) \right)^{-1}
    \left(\sum_{k,l} w^{k,\hat{r}}_{\ell} w^{l,\hat{s}}_{\ell}  \frac{\partial C_{\ell}^{kl}}{\partial A_{\beta}} \right)  \right|_{A_{\alpha}=1=A_{\beta}} \,.
\end{align}
Here we have assumed fixed HILC weights, e.g, as would be determined once from the actual data and applied to every simulation. Thus, the weights are not parameter-dependent quantities (if they were, then the derivatives in Eq.~\eqref{eq.HILCderiv} would be more complicated). Note that there are some subtleties in the indexing. In particular, we must define some mapping from $\hat{p}\hat{q}$ to individual $\hat{p}$ and $\hat{q}$, and similarly for $\hat{r}\hat{s}$. This is discussed in the next subsection.

\subsection{Number of Components vs.~Number of Frequencies}
\label{sec.ncomps_vs_nfreqs}

We prove the equivalence of $F^{\mathrm{mult}}_{\hat{A}_{\alpha} \hat{A}_{\beta}}$ and $F_{\hat{A}_\alpha \hat{A}_\beta}^{\mathrm{HILC}}$ analytically for specific numbers of frequencies and sky components. In all cases, we assume an arbitrary frequency-frequency covariance matrix that is diagonal in $\ell$, unless otherwise noted. The Gaussian covariance matrix as in Refs.~\cite{Louis2017, Kable:2019avl} is an example of such a covariance matrix. Due to the lengthy algebraic expressions in the following analysis, we provide Mathematica notebooks demonstrating each of these results in our GitHub repository \verb|NILC-Inference-Pipeline|.

\subsubsection{2 Components, 2 Frequencies}
This is the simplest case and is also the one we consider in our full simulation-based pipelines. We assume an arbitrary diagonal frequency-frequency power spectrum covariance matrix for this proof, but we do not assume that this covariance matrix is Gaussian. For concreteness, let us assume two frequency channels, e.g., 90 and 150 GHz. Moreover, we assume two sky components, the CMB and tSZ fields, and their respective minimum-variance HILC maps, $\hat{T}$ and $\hat{y}$. In this case, we have that $ij, km \in \{90 \times 90, 90 \times 150, 150 \times 150 \}$. This yields a rank-3 frequency-frequency covariance matrix at each $\ell$, $\mathrm{Cov}^{ij, km}_\ell$. Since there are two components, we have $\hat{p}\hat{q} \in \{\hat{T}\hat{T}, \hat{T}\hat{y}, \hat{y}\hat{y} \}$. The covariance matrix $\mathrm{Cov}^{\hat{p}\hat{q},\hat{r}\hat{s}}_\ell$ has a rank of 3, as can immediately seen via the construction in Eq.~\eqref{eq.HILCcov} in terms of $\mathrm{Cov}^{ij, km}_\ell$.  Thus, the HILC data vector comprises all three spectra $C_\ell^{\hat{T}\hat{T}}, C_\ell^{\hat{T}\hat{y}}, C_\ell^{\hat{y}\hat{y}}$. Using that HILC data vector, the parameter covariance matrix elements from HILC and multifrequency power spectra are equal.

\subsubsection{3 Components, 2 Frequencies}
Let us now suppose that we have three components whose HILC maps are denoted $\hat{T}$, $\hat{y}$, and $\hat{J}$. Suppose we let $C_\ell^{\hat{p}\hat{q}}$ take on values of all the component auto- and cross-spectra, i.e., $C_\ell^{\hat{p}\hat{q}} \in \{C_\ell^{\hat{T}\hat{T}}, C_\ell^{\hat{y}\hat{y}}, C_\ell^{\hat{J}\hat{J}}, C_\ell^{\hat{T}\hat{y}}, C_\ell^{\hat{T}\hat{J}}, C_\ell^{\hat{y}\hat{J}} \}$. We would then have a $6 \times 6$ HILC power spectrum covariance matrix. However, this matrix actually only has a rank of 3 since it is built from the rank 3 frequency-frequency power spectrum covariance matrix, and we must thus remove linearly dependent rows and columns to make the covariance matrix invertible. In particular, the rows and columns corresponding to the elements of the data vector in the  following sets are linearly dependent: $\{C_\ell^{\hat{T}\hat{T}}, C_\ell^{\hat{T}\hat{y}}, C_\ell^{\hat{T}\hat{J}} \}$, $\{C_\ell^{\hat{y}\hat{y}}, C_\ell^{\hat{T}\hat{y}}, C_\ell^{\hat{y}\hat{J}} \}$, and $\{C_\ell^{\hat{J}\hat{J}}, C_\ell^{\hat{T}\hat{J}}, C_\ell^{\hat{y}\hat{J}} \}$.  Using  $C_\ell^{\hat{p}\hat{q}} \in \{C_\ell^{\hat{T}\hat{T}}, C_\ell^{\hat{y}\hat{y}}, C_\ell^{\hat{J}\hat{J}} \}$ as our fundamental HILC data vector thus solves this problem.  Using this construction in Eq.~\eqref{eq.HILC_Fisher_analytic}, we find that $F_{\hat{A}_\alpha \hat{A}_\beta}^{\mathrm{HILC}} = F^{\mathrm{mult}}_{\hat{A}_{\alpha} \hat{A}_{\beta}}$, and thus the HILC and multifrequency-power-spectrum methods yield identical parameter covariance matrix elements.  (The analytic demonstration can be found in the Mathematica notebooks in \verb|NILC-Inference-Pipeline|).

\subsubsection{2 Components, 3 Frequencies}
For concreteness, suppose that the three frequency channels are 90, 150, and 220 GHz. Then there are 6 possible values values for each of $ij$ and $km$: $ij,km \in \{90 \times 90, 150 \times 150, 220 \times 220, 90 \times 150, 90 \times 220, 150 \times 220 \}$. The frequency-frequency power spectrum covariance matrix thus has a rank of 6. Since there are two components in the sky model (let us call their associated HILC maps $\hat{T}$ and $\hat{y}$), the largest possible HILC data vector is $\{ C_\ell^{\hat{T}\hat{T}}, C_\ell^{\hat{T}\hat{y}}, C_\ell^{\hat{y}\hat{y}} \}$. This would result in a HILC power spectrum covariance matrix of rank 3, which is less than the rank of the frequency-frequency power spectrum covariance matrix. As a result, the two methods do not yield identical parameter covariance matrix elements in this scenario. Analytically, it is not obvious which one has a larger covariance, but numerically, we find that the parameter covariance matrices are actually nearly identical in practice (see Appendix \ref{app:3freqs}).

\subsubsection{3 Components, 3 Frequencies}
Suppose that the three frequency channels are 90, 150, and 220 GHz. Then there are 6 possible values values for each of $ij$ and $km$: $ij,km \in \{90 \times 90, 150 \times 150, 220 \times 220, 90 \times 150, 90 \times 220, 150 \times 220 \}$. The frequency-frequency power spectrum covariance matrix thus has a rank of 6. Let the HILC maps of the three sky components be denoted as $\hat{T}$, $\hat{y}$, and $\hat{J}$. Then $\hat{p}\hat{q}$ may take on values of all the auto- and cross-spectra, i.e., $\hat{p}\hat{q} \in \{C_\ell^{\hat{T}\hat{T}}, C_\ell^{\hat{y}\hat{y}}, C_\ell^{\hat{J}\hat{J}}, C_\ell^{\hat{T}\hat{y}}, C_\ell^{\hat{T}\hat{J}}, C_\ell^{\hat{y}\hat{J}} \}$. This matrix also has a rank of 6 since it is built from the frequency-frequency covariance matrix of rank 6. Thus, in this case we can use all the possible auto- and cross-spectra in the HILC data vector. Due to computational constraints of running this scenario entirely analytically, we assume a Gaussian frequency-frequency power spectrum covariance matrix and derive analytic results for the final Fisher matrices in both the HILC and multifrequency power spectrum approaches. Testing several numerical set-ups, we then find that the HILC and multifrequency approaches yield identical parameter covariance matrices, as in the two-component, two-frequency case considered above.

\subsubsection{General Conclusions}
Based on these simple examples, we can extrapolate some general conclusions:
\begin{itemize}
    \item  When the number of frequencies is equal to the number of components, we can use all possible HILC map auto- and cross-spectra in the data vector for the HILC approach, and the two approaches yield identical results for the final parameter covariance matrix.
    \item When there are more components than frequencies, the rank of the HILC power spectrum covariance matrix is limited by the rank of the frequency-frequency power spectrum covariance matrix, and we must thus limit the number of spectra in the HILC data vector. After doing so, we find that the two approaches also yield equal parameter covariance 
    matrices.
    \item When there are more frequencies than components, the rank of the HILC power spectrum covariance matrix is necessarily less than that of the frequency-frequency power spectrum covariance matrix. Thus, the two methods do not give equivalent results analytically, though numerically we find that they do give nearly identical results, with the HILC posterior often being very slightly larger (see Appendix \ref{app:3freqs}). 
\end{itemize}

\section{Comparing Posterior Distributions from Simulations}
\label{sec.results}

\begin{figure}[t]
    \centering
    {\includegraphics[width=0.49\textwidth]{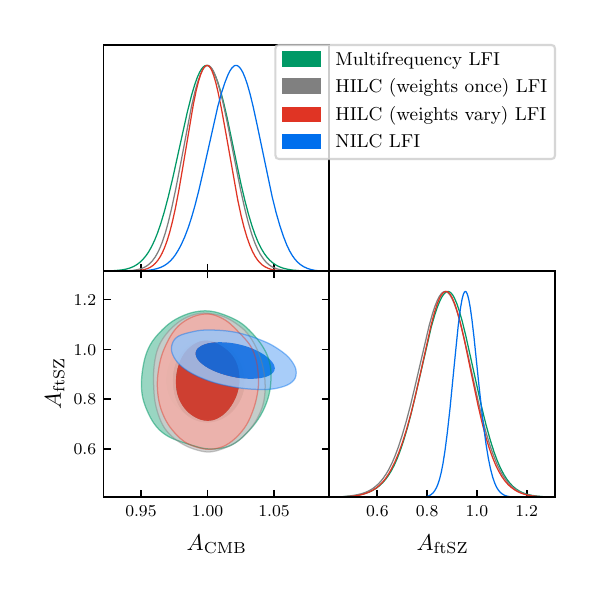}}
    {\includegraphics[width=0.49\textwidth]{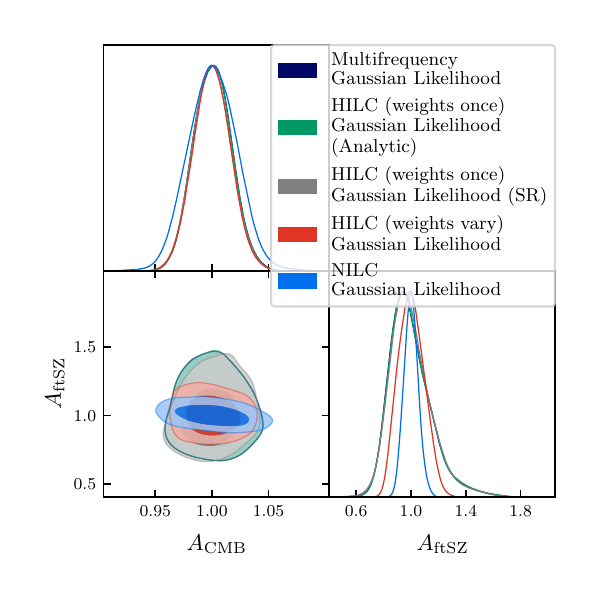}}
    \caption{Parameter posteriors obtained via LFI (left) and Gaussian likelihoods (right). In both figures, we compare the results using various summary statistics: multifrequency power spectra (described in Sec.~\ref{sec.method_multifrequency_power_spec}), HILC power spectra (with a few variants based on weight determination and parameter dependence determination that are described in Sec.~\ref{sec.method_hilc}), and NILC power spectra (described in Sec.~\ref{sec.method_nilc}). For LFI, we use 40000 simulations for each pipeline. For the Gaussian likelihood, we use 2000 simulations and show the distribution of maximum-likelihood estimates from each simulation. It is evident that the area of the 2D posterior is significantly smaller in the NILC pipeline than the traditional multifrequency pipeline, suggesting that NILC is a useful transformation of the data to use in parameter inference pipelines when there are non-Gaussian sky components. }
    \label{fig:final_posteriors}
\end{figure}

\begin{figure}[t]
    \centering
    {\includegraphics[width=0.4\textwidth]{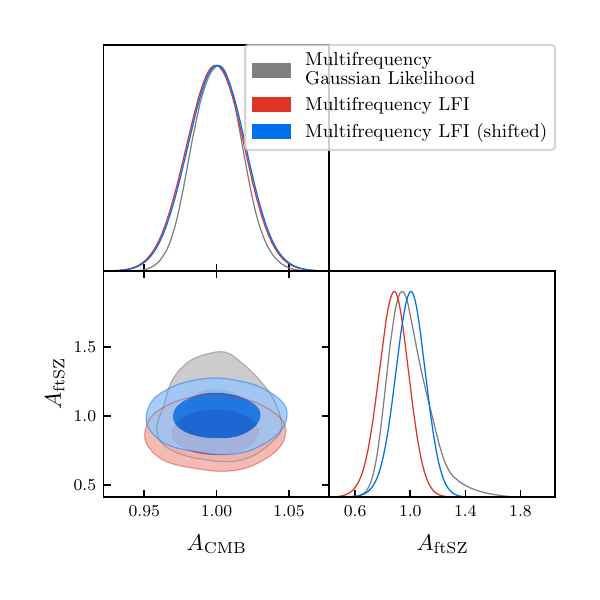}}
    {\includegraphics[width=0.4\textwidth]{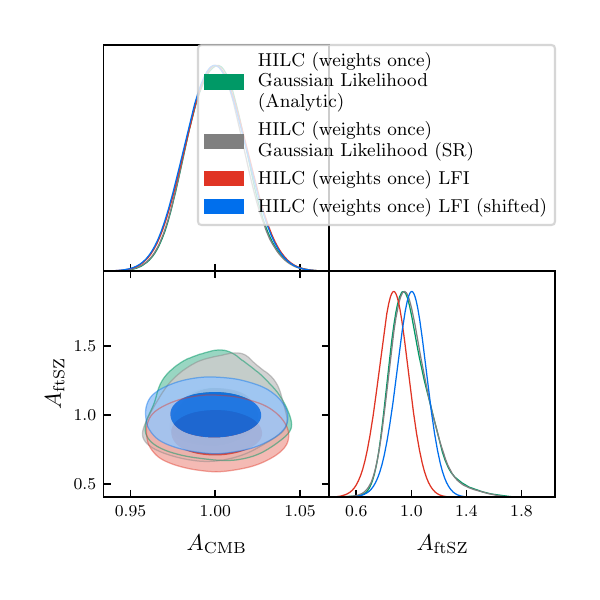}}
    {\includegraphics[width=0.4\textwidth]{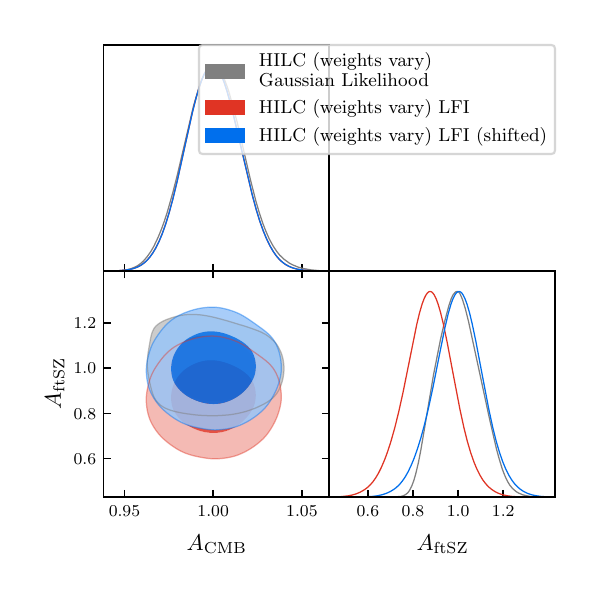}}
    {\includegraphics[width=0.4\textwidth]{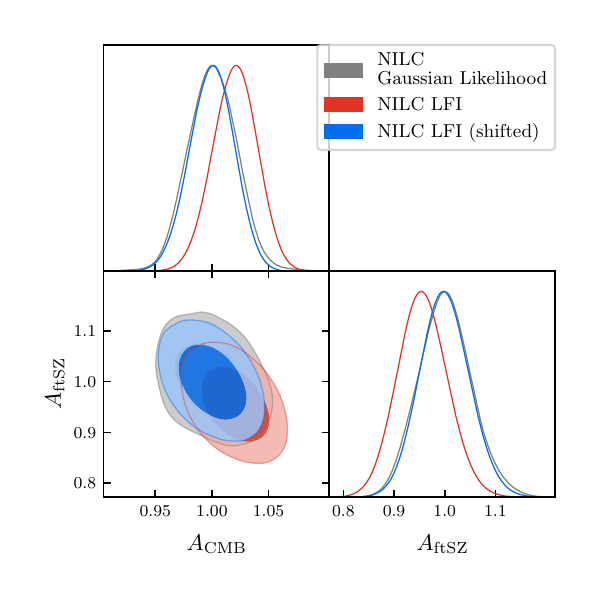}}
    \caption{Comparison of posteriors from LFI to those from a Gaussian likelihood for all the summary statistics considered in this work. We show posteriors for the Gaussian likelihood, LFI, and LFI shifted such that the posterior is centered on 1.0 for both parameters. The Gaussian likelihood is inaccurate in the presence of the highly amplified non-Gaussian tSZ field.}
    \label{fig:lfi_vs_gaussianlkl}
\end{figure}

In this section, we compare the simulation-derived posteriors obtained from the multifrequency power spectrum, harmonic ILC, and needlet ILC approaches. The posterior distributions for both LFI and the Gaussian likelihood (with posteriors computed via MLE here) are shown in Fig.~\ref{fig:final_posteriors}. In both cases, we see that NILC significantly shrinks the area of the 2D posterior compared to the other approaches. In particular, we see an approximately $60\%$ reduction in the area of the 68\% confidence interval 2D posterior (in the elliptical approximation, discussed further in Appendix \ref{app:3freqs}) using NILC power spectra as compared with multifrequency power spectra as summary statistics for LFI. Fig.~\ref{fig:lfi_vs_gaussianlkl} compares the posteriors from LFI and the Gaussian likelihood for each of the pipelines to assess whether the Gaussian likelihood is a valid assumption. We predict that it is not a good assumption since, if we histogram the value of one summary statistic in one bin from several simulations, the histogram is non-Gaussian. From the changes in the posteriors (computed in the Gaussian likelihood versus with LFI), we see that it is indeed an imperfect assumption. Table \ref{table:posteriors} shows the numerical values of 1D marginalized posteriors for all the methods considered in this work.

As a consistency check, we also run all of the pipelines in this work using a set-up containing only Gaussian random fields (i.e., we generate Gaussian realizations of the ftSZ field using a fiducial power spectrum). In this situation, there should be no advantage to performing NILC over HILC, and moreover, the Gaussian likelihood should yield the same results as LFI.\footnote{When data are cosmic-variance-limited, LFI is more optimal and less biased than the Gaussian likelihood, even when every field in the problem is Gaussian. This is because the covariance matrix in the Gaussian likelihood is actually a parameter-dependent quantity, but generally treated as parameter-independent, while LFI can learn such a dependence \cite{vonWietersheim-Kramsta:2024cks}. However, in the situation studied here, there is non-negligible instrumental noise, so the data are not cosmic-variance-limited, and thus we expect the two approaches to yield nearly identical results.} This is explored in Appendix \ref{app:gaussian_random_fields}, where we find that this is indeed the case, and the posteriors from all methods are nearly identical.

\begin{table}[t]
\setlength{\tabcolsep}{10pt}
\renewcommand{\arraystretch}{1.3}
    \begin{tabular}{|l|l|l|}
        \hline
         \textbf{Pipeline (on non-Gaussian simulations)} &  $A_{\mathrm{CMB}}$ & $ A_{\mathrm{ftSZ}}$  \\
         \hline \hline 
         \textbf{Multifrequency PS} &MLE: $1.001 \pm 0.017 $    &MLE: $0.999^{+0.10}_{-0.18}$   \\
          -Gaussian likelihood &Fisher: \_\_ $\pm 0.017$    &Fisher: \_\_ $\pm 0.16$   \\
          -analytic parameter dependence &MCMC: $1.000\pm 0.016$    &MCMC: $ 0.99\pm 0.16$   \\
           
         \hline
          \textbf{Multifrequency PS} &$0.999 \pm 0.020$  &$0.88 \pm 0.11$  \\
          -likelihood-free inference &  &  \\
        \hline \hline

        \textbf{Harmonic ILC} &MLE: $1.001 \pm 0.017$    &MLE: $0.999^{+0.10}_{-0.18}$   \\
        -fixed weights &Fisher: \_\_ $\pm 0.017$    &Fisher: \_\_ $\pm 0.16$   \\
        -Gaussian likelihood &MCMC: $1.000^{+0.017}_{-0.015}$    &MCMC: $1.01\pm 0.16$   \\
        -analytic parameter dependence &    &   \\
        \hline

        \textbf{Harmonic ILC} &MLE: $1.000 \pm 0.017$  &MLE: $1.00^{+0.10}_{-0.17}$   \\
        -fixed weights &Fisher: \_\_ $\pm 0.017$     &Fisher: \_\_ $\pm 0.16$   \\
        -Gaussian likelihood &MCMC: $ 0.9996\pm 0.017$   &MCMC: $0.999\pm 0.16$   \\
        -parameter dependence from symbolic regression &   &   \\
        \hline
        
        \textbf{Harmonic ILC} &$1.001 \pm 0.017$    &$0.87 \pm 0.11$   \\
        -fixed weights &    &   \\
        -likelihood-free inference &    &   \\
        \hline

        \textbf{Harmonic ILC} &MLE: $1.000 \pm 0.016$    &MLE: $1.001^{+0.089}_{-0.10}$   \\
        -varying weights &Fisher: \_\_ $\pm 0.016$    &Fisher: \_\_ $\pm 0.11$   \\
        -Gaussian likelihood &MCMC: $1.002\pm 0.016$    &MCMC: $1.035^{+0.073}_{-0.11}$   \\
        -parameter dependence from symbolic regression &    &   \\
        \hline
        
        \textbf{Harmonic ILC} &$1.000 \pm 0.015$    &$0.87 \pm 0.11$   \\
        -varying weights &    &   \\
        -likelihood-free inference &    &   \\
        \hline
        \hline

        \textbf{Needlet ILC} &MLE: $1.001 \pm 0.022$    &MLE: $1.001 \pm 0.055$   \\
        -Gaussian likelihood &Fisher: \_\_ $\pm 0.022$   &Fisher: \_\_ $\pm 0.052$   \\
        -parameter dependence from symbolic regression &MCMC: $1.001\pm 0.022$    &MCMC: $1.007^{+0.049}_{-0.054}$   \\
        \hline
        
        \textbf{Needlet ILC} &$1.020 \pm 0.019$ &$0.956 \pm 0.049$   \\
        -likelihood-free inference &  &   \\
        \hline
        
    \end{tabular}
    \caption{Parameter constraints for all the methods considered in this work. See Sec.~\ref{sec.method_multifrequency_power_spec}, \ref{sec.method_hilc}, and \ref{sec.method_nilc} for information on using multifrequency power spectra, harmonic ILC power spectra, and needlet ILC power spectra as summary statistics, respectively. We present 1D marginalized posterior values obtained with the different summary statistics using both likelihood-free inference and a Gaussian likelihood. For the Gaussian likelihood, we compute posteriors using maximum-likelihood estimation (MLE), Fisher matrix calculations, and a Markov chain Monte Carlo (MCMC) algorithm. Note that we do not quote a central value for the Fisher matrix calculation since it only predicts the error bars (which are assumed to be symmetric in the Fisher approximation).}
    \label{table:posteriors}
\end{table}

As discussed in Sec.~\ref{sec.ncomps_vs_nfreqs}, when the number of frequencies is equal to the number of components in the sky model (as is the case here), the HILC fixed-weights pipeline and multifrequency-power-spectra pipeline must yield identical results for the Gaussian-likelihood analysis. We have shown that when the number of components is greater than the number of frequency channels, the fixed-weights HILC power spectrum covariance matrix becomes singular if we make HILC maps for every component in the sky model and use all of their auto- and cross-power spectra in the data vector. A natural question is whether the same would be true for NILC power spectra, which have nontrivial scale-dependent coupling. Using our simulations, we show that, to within our numerical precision, the determinant of the NILC power spectrum covariance matrix is zero in such a situation. Specifically, we demonstrate this on a set-up very similar to the main one used in the simulations, but with an added Gaussian realization of the CIB field, assuming a modified blackbody SED \cite{Madhavacheril_2020} with an effective dust temperature of 20.0~K and spectral index of 1.45. With this set-up, we also verify numerically that the HILC power spectrum covariance matrices are singular for both variants of the HILC pipeline (fixed weights and varying weights). The implication is that, for both the HILC and NILC inference approaches, if the number of components is greater than the number of frequency channels, one must carefully construct the fundamental data vector such that the elements are linearly independent (i.e. one should not use the auto- and cross-spectra of every component ILC map in the data vector). 

We also examine how the results change when varying the frequency channels used or when increasing the number of frequency channels to three (while still keeping just two components). These results are shown in Appendix \ref{app:3freqs}. In all cases, using NILC power spectra as summary statistics either shrinks or yields the same area of the 2D posterior as compared with using multifrequency power spectra as summary statistics.  This method thus has significant promise in yielding tighter parameter constraints from ongoing and upcoming CMB experiments.


\section{Discussion}
\label{sec.discussion}

In this paper we have considered several methods for inferring parameters describing the power spectra of physical sky components from multifrequency mm-wave data.  We have shown that using using auto- and cross-spectra of HILC maps in a Gaussian likelihood yields the same final parameter error bars as the traditional approach of fitting templates to multifrequency auto- and cross-power spectra, with caveats of which HILC spectra to include in the data vector based on the number of frequency channels and number of sky components (discussed in Sec.~\ref{sec.method_hilc_gaussian_lkl_proof}). Specifically, when the number of sky components is greater than or equal to the number of frequency channels, we have analytically shown that performing inference with harmonic ILC power spectra (with fixed weights) yields identical parameter covariance matrices as performing inference with multifrequency power spectra on Gaussian random fields. However, the power spectra of HILC maps only contain Gaussian information, and many mm-wave foreground fields vary spatially on the sky and are non-Gaussian. Thus, NILC is a better candidate to capture such additional information. In previous work, we derived an analytic expression for the auto- and cross-spectra of NILC maps in Ref.~\cite{surrao2024_nilc1}, finding a nontrivial parameter dependence. In this work, we have estimated this parameter dependence from simulations using symbolic regression, and have then performed a maximum-likelihood analysis to show that NILC reduces the area of the 68\% confidence interval 2D posterior on CMB and ftSZ amplitude parameters by 60\%, with results summarized in Table \ref{table:posteriors} and 2D and 1D marginalized posterior distributions shown in Fig.~\ref{fig:final_posteriors}.

Because the weight maps are functions of the fields, the results of Ref.~\cite{surrao2024_nilc1} demonstrate that the parameter dependence of NILC power spectra is highly non-trivial. With NILC, the exact dependence of the weight maps on the fields is complex, so we must estimate this dependence using simulations and symbolic regression to avoid assuming a specific functional form. Fig.~\ref{fig:SR} in Appendix \ref{app:sr} shows the complexity of this dependence, which depends on the exact amplitude of each component. In some cases, we note that when the amplitude of some component increases, its contribution to the NILC power spectrum of another component actually decreases since the weight maps then prioritize suppressing this contaminant over the others. With such complex parameter dependence, and given that in many situations the Gaussian likelihood is an imperfect assumption, LFI is a much simpler and more accurate approach for determining parameter posteriors in the presence of non-Gaussian contaminants. Similar to the case when using the Gaussian likelihood, we find that performing LFI with NILC power spectra as summary statistics reduces the area of the 68\% confidence interval 2D posterior on CMB and ftSZ amplitude parameters by 60\% as compared with performing LFI with multifrequency power spectra as summary statistics. 

In Appendix \ref{app:3freqs} we examine various frequency combinations and how they impact the results of the different summary statistics. In all cases, NILC either maintains or shrinks the area of the 2D posterior, showing that NILC component map power spectra are powerful summary statistics. In our set-up, we have used 40,000 simulations in the LFI, using a single round of NPE with a uniform prior. To be more efficient in the number of simulations, one could use sequential neural posterior estimation (SNPE), running the NPE procedure in multiple rounds \cite{SNPE1, SNPE2, SNPE3, SNPE4}. In the first round, parameters are sampled from the prior and used to generate simulations, learning an initial posterior. In subsequent rounds, parameters are sampled instead from the current learned posterior rather than the prior. Another modification to be more efficient in the number of simulations is to use a very wide Gaussian prior rather than a uniform prior.

Here we have demonstrated our result on a somewhat contrived scenario in which the tSZ field is amplified by a factor of 150, but we would expect similar trends in the error bars with the usual tSZ field at high $\ell$, in particular, beyond $\ell=2000$ where the tSZ signal amplitude surpasses that of the CMB at 90 GHz. Moreover, we would expect such large error bar reductions in other cases where non-Gaussian foregrounds dominate over the signal of interest, as is the case when searching for primordial \textit{B}-modes, for which the non-Gaussian dust dominates. With the latest constraint on the tensor-to-scalar ratio of $r_{0.05}< 0.036$ \cite{BICEP:2021xfz}, the dust amplitude is approximately a factor of 10-$100$ times larger than the CMB \textit{B}-mode amplitude at 150 GHz for $2 \leq \ell \leq 500$ (see Fig.~12 of Ref.~\cite{Planck:2018gnk}), depending on the region of sky analyzed. This ratio is even larger than the ratio in our toy model of the ftSZ to CMB (which is within a factor of a few, cf.~Fig.~\ref{fig:input_spectra}). The current limitation in demonstrating our NILC approach on the example of primordial \textit{B}-modes and polarized dust is the need for many realistic anisotropic, non-Gaussian simulations of the dust field.  However, progress has been made toward such simulations in recent work \cite{Krachmalnicoff:2020, Thorne:2021, Regaldo-SaintBlancard:2020dlb}, and is also the focus of follow-up work in preparation.

This work demonstrates the ability of NILC to capture non-Gaussian information, as well as the constraining power that such information can have on parameters. It thus motivates further methodological development for using NILC-based inference or other techniques that efficiently capture information beyond two-point statistics in analyses of the mm-wave sky.

\section{Acknowledgments}
We thank Marcelo Alvarez for help with generating tSZ simulations with \verb|halosky|; and Adrian Bayer, Sam Goldstein, Mathew Madhavacheril, Fiona McCarthy, Chirag Modi, Shivam Pandey, Oliver Philcox, and Kendrick Smith for useful discussions. We also thank Adri Duivenvoorden for comments on the manuscript. This material is based upon work supported by the National Science Foundation Graduate Research Fellowship Program under Grant No. DGE 2036197 (KMS). JCH acknowledges support from NSF grant AST-2108536, NASA grant 80NSSC22K0721, NASA grant 80NSSC23K0463, DOE grant DE-SC00233966, the Sloan Foundation, and the Simons Foundation. Several software tools were used in the development and presentation of results shown in this paper, including \verb|getdist| \cite{getdist}, \verb|HEALPix/healpy| \cite{Healpix, Healpy}, \verb|numpy| \cite{numpy}, \verb|scipy| \cite{scipy}, \verb|matplotlib| \cite{matplotlib}, \verb|astropy| \cite{astropy1, astropy2, astropy3}, \verb|emcee| \cite{emcee}, \verb|PySR| \cite{pysr}, \verb|sbi| \cite{sbi}, \verb|pyilc| \cite{McCarthy:2023hpa, McCarthy:2023cwg}, \verb|wandb| \cite{wandb}, \verb|PyTorch| \cite{pytorch}, and \verb|Julia| \cite{Julia-2017}. The authors acknowledge the Texas Advanced Computing Center (TACC) at The University of Texas at Austin for providing HPC resources that have contributed to the research results reported within this paper. The authors also acknowledge the use of resources of the National Energy Research Scientific Computing Center (NERSC), a U.S.~Department of Energy Office of Science User Facility located at Lawrence Berkeley National Laboratory. Finally, the authors acknowledge the use of computing resources from Columbia University's Shared Research Computing Facility project, which is supported by NIH Research Facility Improvement Grant 1G20RR030893-01, and associated funds from the New York State Empire State Development, Division of Science Technology and Innovation (NYSTAR) Contract C090171, both awarded April 15, 2010. 

\begin{appendices}


\section{ILC Maps and Weight Maps}
\label{app:maps}

\begin{figure}[htb]
    \centering
    \includegraphics[width=0.5\textwidth]{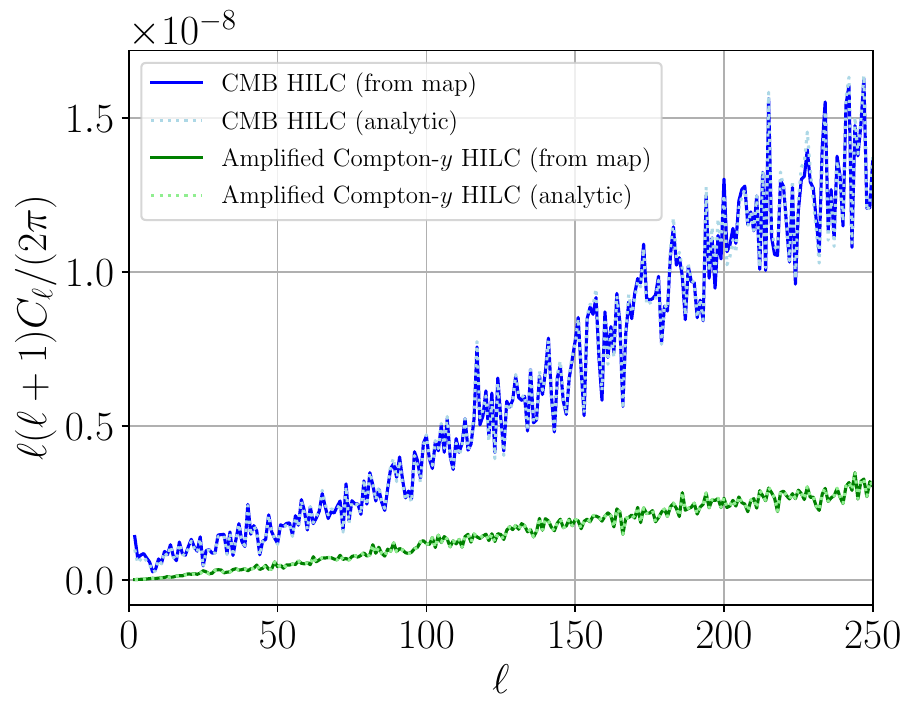}
    \caption{Comparison of harmonic ILC spectra computed analytically as described in Sec.~\ref{sec.method_hilc} (dotted curves) and spectra computed from a map-based HILC operation using \texttt{pyilc} (solid curves) for a single simulation. The curves are shown in units of $\mathrm{K}^2$ for the CMB spectra and dimensionless Compton-$y$ units for the Compton-$y$ spectra. Harmonic ILC weights are computed with $\Delta \ell=20$ (see Sec.~\ref{sec.method_hilc}). There is no $\ell$-space binning shown in the plot.}
    \label{fig:hilc_map_vs_analytic}
\end{figure}

\begin{figure}[htb]
    \centering
    {\includegraphics[width=0.9\textwidth]{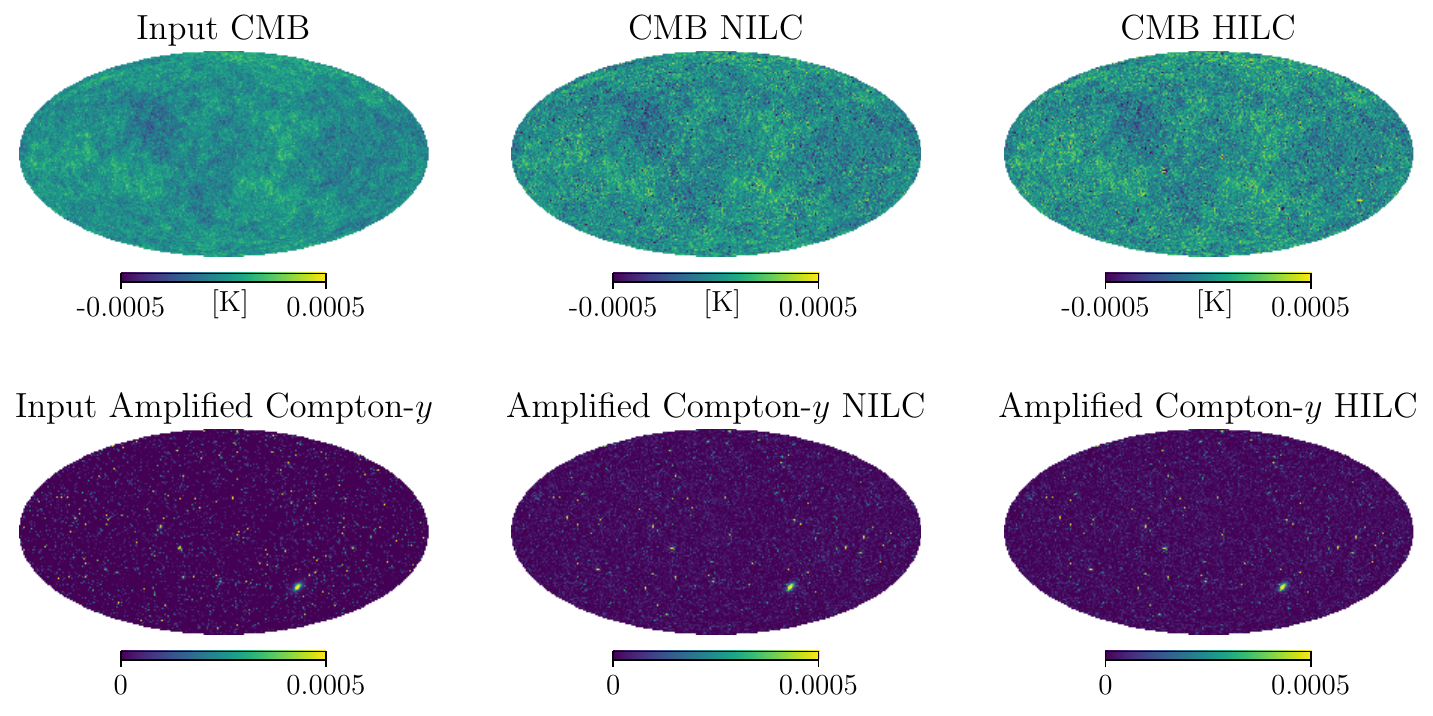}}
    \caption{Comparison of input CMB and amplified Compton-$y$ maps to the maps reconstructed via NILC and HILC. Residual bright spots resulting from ftSZ contamination are present in the CMB HILC map but not in the CMB NILC map, demonstrating the ability of NILC to better clean non-Gaussian contaminants.} 
    \label{fig:recon_maps}
\end{figure}

\begin{figure}[htb]
    \centering
    {\includegraphics[width=0.49\textwidth]{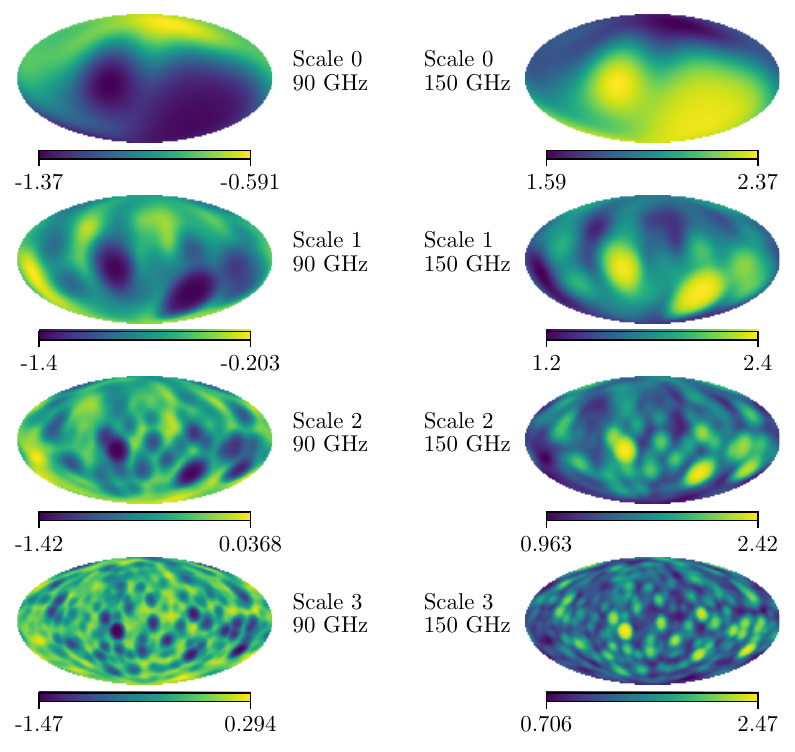}}
    {\includegraphics[width=0.49\textwidth]{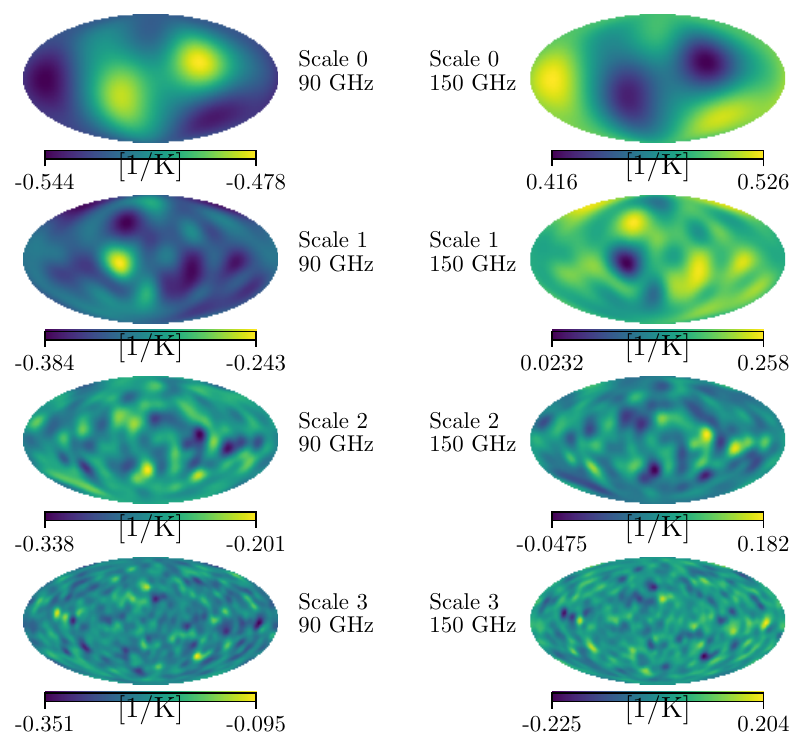}}
    \caption{NILC weight maps for a preserved CMB component (left two columns) and preserved tSZ component (right two columns), generated with \texttt{pyilc}, for a sky model consisting of the CMB, amplified tSZ signal (ftSZ), and noise. The maps are shown for 90 GHz and 150 GHz at each needlet filter scale, with scale 0 corresponding to the lowest-$\ell$ scale and scale 3 corresponding to the highest-$\ell$ scale. The CMB and ftSZ maps underlying the frequency maps from which these NILC weight maps are produced are shown in Fig.~\ref{fig:recon_maps}. The CMB weight maps are unitless, and the Compton-$y$ weight maps have units of 1/K such that the output NILC map is in dimensionless Compton-$y$ units.  }
    \label{fig:wt_maps}
\end{figure}

In this appendix we show images of component maps and ILC weight maps. For the harmonic ILC pipeline, we compute the spectra that comprise the data vector analytically, as described in Sec.~\ref{sec.method_hilc}. However, one can also explicitly construct a HILC map and then compute the power spectrum of that map. In Fig.~\ref{fig:hilc_map_vs_analytic}, we compare the analytic spectra to the map-based spectra, confirming that the results are effectively identical.

In Fig.~\ref{fig:recon_maps}, we compare the reconstructed CMB and amplified Compton-$y$ maps from the NILC and HILC pipelines. The reconstructed amplified Compton-$y$ maps look very similar. However, in the reconstructed CMB maps one can see some bright spots in the HILC map that are not present in the input CMB map or NILC CMB map. These coincide with bright spots in the ftSZ map. Thus, this is an example of how HILC does not optimally clean out non-Gaussian contaminants, but NILC does.

Fig.~\ref{fig:wt_maps} shows the NILC weight maps used for the construction of both the CMB-preserved NILC map and ftSZ-preserved NILC map. These weight maps are built from the same map realizations shown in Fig.~\ref{fig:recon_maps}. We see that the weight maps have the general appearance of the contaminant field since the purpose of the weight maps is to suppress contamination. For example, the weight maps for a CMB-preserved NILC map roughly mimic the structure of the ftSZ map in Fig.~\ref{fig:recon_maps}.


\section{Parameter Dependence from Symbolic Regression}
\label{app:sr}

\begin{figure}[htb]
    \centering
    {\includegraphics[width=0.7\textwidth]{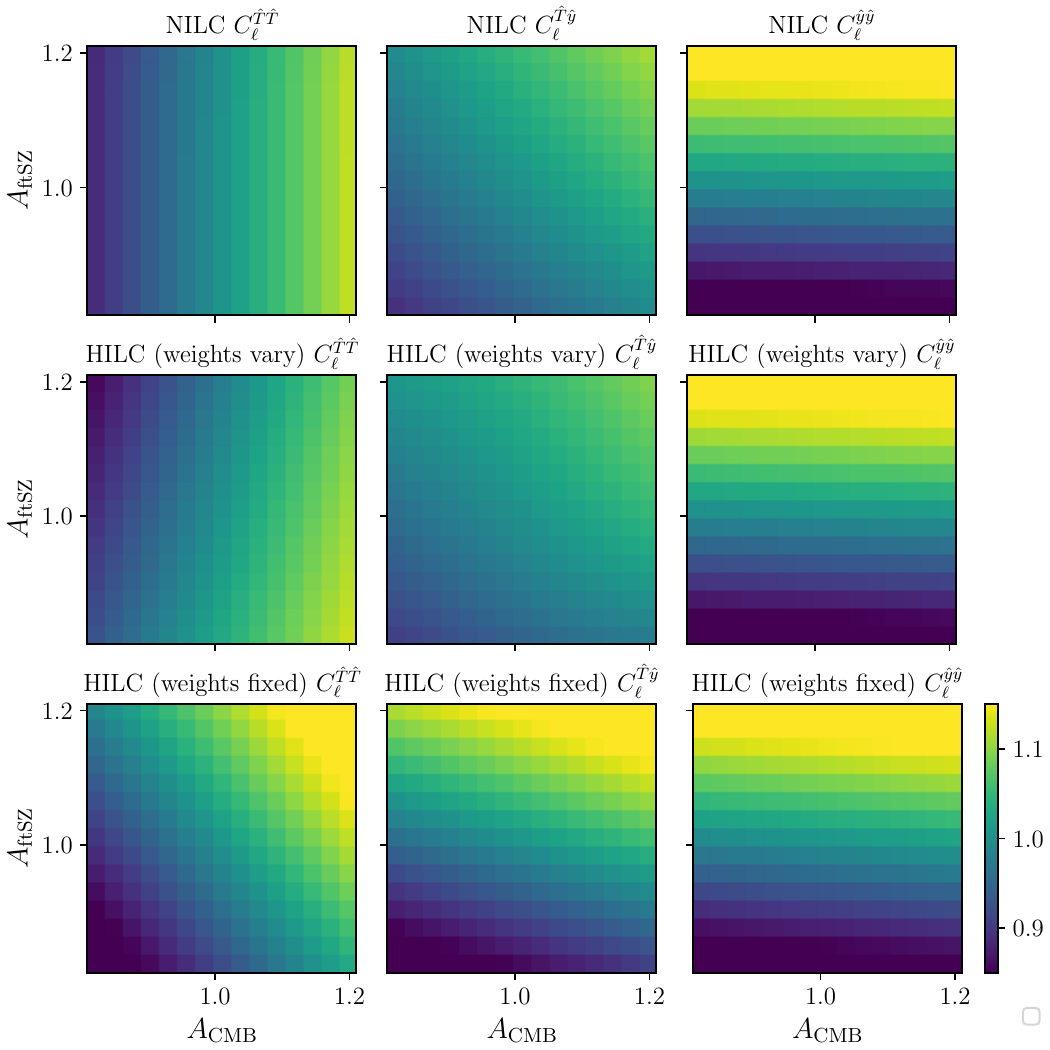}}
    \caption{Visual representation of the parameter dependence of ILC component map power spectra obtained via symbolic regression for NILC power spectra (top row), HILC power spectra with weights determined separately for each simulation (middle row), and HILC power spectra with weights determined from a fiducial template and then applied to every simulation (bottom row). Parameter dependence is shown for the different elements of the data vector: $C_\ell^{\hat{T}\hat{T}}$ (left), $C_\ell^{\hat{T}\hat{y}}$ (middle), and $C_\ell^{\hat{y}\hat{y}}$ (right). In particular, we plot $f^{\hat{p}\hat{q}}_{b}(A_{\mathrm{CMB}}, A_{\mathrm{ftSZ}})$ from Eq.~\eqref{eq.sr_f_def}, where $b=5$ here, corresponding to a mean multipole $\ell=138$.  }
    \label{fig:SR}
\end{figure}

In this appendix we discuss results for determining the parameter dependence of HILC and NILC power spectra via symbolic regression. The details of this procedure are presented in Sec.~\ref{sec.method_hilc_gaussian_lkl_sr}. In Fig.~\ref{fig:SR} we show the values of $f^{\hat{p}\hat{q}}_{b}(A_{\mathrm{CMB}}, A_{\mathrm{ftSZ}})$ from Eq.~\eqref{eq.sr_f_def} for the different $\hat{p}, \hat{q}$ and different methods considered in this work. Since $f_b$ is a separate function for each bin, we choose one bin for which to plot the results (here it is $b=5$, which corresponds to a mean multipole of $\ell=138$). We show the values of $f_b$ for the region of parameter space around the fiducial $A_{\mathrm{CMB}}=1=A_{\mathrm{ftSZ}}$.

We also show the analytic expressions output by the regressor for bin 5 in Table~\ref{tab:SR}. From these expressions, we see that the regressor may be slightly overfitting (for example, this is apparent from the term involving $A_{\mathrm{ftSZ}}$ in the expression for NILC $C_b^{\hat{T}\hat{T}}$), but this is not important for our purposes here. We simply require expressions that give smooth and accurate parameter dependence around the fiducial $A_{\mathrm{CMB}}=1=A_{\mathrm{ftSZ}}$.

\begin{table}[htb]
\setlength{\tabcolsep}{10pt}
\renewcommand{\arraystretch}{2.0}
    \begin{tabular}{|c|c|}
    \hline 
    \textbf{Spectrum} & \textbf{Expression} \\
    \hline
      NILC $C_{b=5}^{\hat{T}\hat{T}}$  & $0.57870084A_{\mathrm{CMB}} - 1.53680702801541 \times 10^{-5} (A_{\mathrm{ftSZ}})^{27} + 0.42053628$  \\
      \hline
    NILC $C_{b=5}^{\hat{T}\hat{y}}$  & $\mathrm{exp}(\frac{A_{\mathrm{CMB}} - 1/A_{\mathrm{ftSZ}}}{A_{\mathrm{CMB}}/A_{\mathrm{ftSZ}} + 2.737616})$  \\
      \hline
    NILC $C_{b=5}^{\hat{y}\hat{y}}$  & $0.03574537 A_{\mathrm{CMB}} - 0.03574537(A_{\mathrm{ftSZ}})^2 + A_{\mathrm{ftSZ}}$  \\
      \hline
      HILC (weights vary) $C_{b=5}^{\hat{T}\hat{T}}$  & $\mathrm{exp} \{[(A_{\mathrm{CMB}})^4 - A_{\mathrm{ftSZ}}] \times \mathrm{exp}(-2A_{\mathrm{CMB}}) \}$      \\
      \hline
    HILC (weights vary) $C_{b=5}^{\hat{T}\hat{y}}$  & $\mathrm{exp}\{0.22931802A_{\mathrm{CMB}} - 0.22931802 \times \frac{\mathrm{exp}[(A_{\mathrm{CMB}} - A_{\mathrm{ftSZ}})^3]}{A_{\mathrm{ftSZ}}}\}$      \\
      \hline
    HILC (weights vary) $C_{b=5}^{\hat{y}\hat{y}}$  & $0.036737457A_{\mathrm{CMB}} - 0.036737457(A_{\mathrm{ftSZ}})^2 + A_{\mathrm{ftSZ}}$     \\
      \hline
      HILC (weights fixed) $C_{b=5}^{\hat{T}\hat{T}}$ & $0.5588486A_{\mathrm{CMB}} + 0.4411514A_{\mathrm{ftSZ}}$  \\
      \hline
    HILC (weights fixed) $C_{b=5}^{\hat{T}\hat{y}}$ & $0.24519494A_{\mathrm{CMB}} + 0.75480506A_{\mathrm{ftSZ}}$ \\
      \hline
    HILC (weights fixed) $C_{b=5}^{\hat{y}\hat{y}}$ & $0.076038094388331A_{\mathrm{CMB}} + 0.923961905611669A_{\mathrm{ftSZ}}$ \\
      \hline
    \end{tabular}
    \caption{Parameter dependence of ILC power spectra obtained via symbolic regression for NILC power spectra, HILC power spectra with weights determined separately for each simulation, and HILC power spectra with weights determined from a fiducial template and then applied to every simulation. Parameter dependence is shown for the different elements of the data vector: $C_\ell^{\hat{T}\hat{T}}$, $C_\ell^{\hat{T}\hat{y}}$, and $C_\ell^{\hat{y}\hat{y}}$. In particular, we show the resulting expressions for $f^{\hat{p}\hat{q}}_{b}(A_{\mathrm{CMB}}, A_{\mathrm{ftSZ}})$ from Eq.~\eqref{eq.sr_f_def}, where $b=5$ here, corresponding to a mean multipole $\ell=138$.}
    \label{tab:SR}
\end{table}


\section{Results on Gaussian Random Fields}
\label{app:gaussian_random_fields}

In this appendix we present the same results as in Sec.~\ref{sec.results}, but evaluated on simulations containing only Gaussian random fields. The CMB and noise remain the same as described in Sec.~\ref{sec.methods}, but the ftSZ field is now a Gaussian random field as well. Specifically, we take the power spectrum of one tSZ map generated via \verb|halosky|, where the map is amplified by a factor of 150. Then for each simulation, we generate a Gaussian realization of that power spectrum. With all Gaussian random fields, we expect that the posterior distributions should be nearly identical for all of the methods considered in this work. In particular, the Gaussian likelihood should be equivalent to likelihood-free inference since the Gaussian likelihood assumption must hold when every field in the problem is Gaussian random.\footnote{Strictly speaking, this is only true for sufficiently high $\ell$ --- at low $\ell$ there can still be deviations from Gaussianity in the likelihood due to the small number of harmonic modes, as described in, e.g., Ref.~\cite{Hamimeche:2008ai}.} Moreover, without the presence of large non-Gaussian fields, we would expect that there is no additional information for NILC to extract that is not already encompassed by the two-point functions contained in the multifrequency power spectra. Thus, we also expect that using multifrequency power spectra, HILC power spectra, and NILC power spectra as summary statistics should all yield nearly identical results. We use this fact to validate all of the methods presented in this paper and to assess the significance of 2D posterior area reductions in the full non-Gaussian ftSZ field case.

\begin{figure}[htb]
    \centering
    {\includegraphics[width=0.49\textwidth]{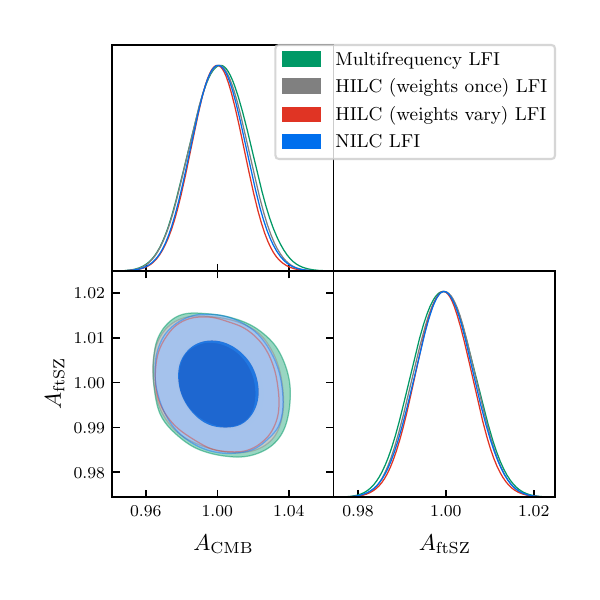}}
    {\includegraphics[width=0.49\textwidth]{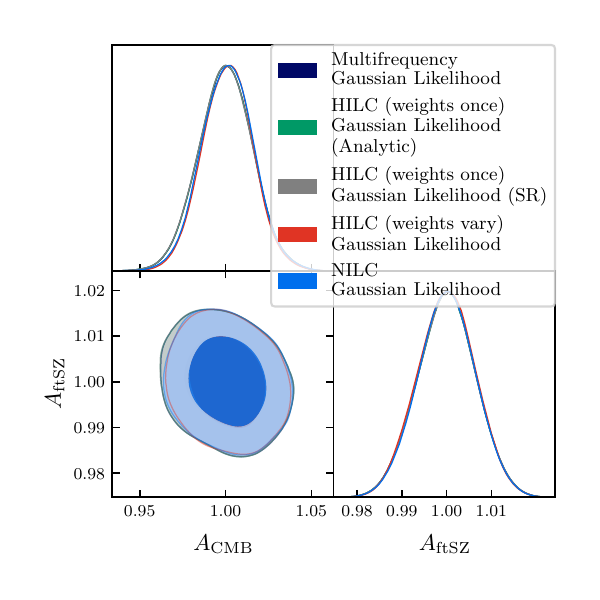}}
    \caption{Parameter posteriors obtained via LFI (left) and Gaussian likelihoods (right) using simulations containing only Gaussian random fields (here the ``ftSZ" field is a Gaussian realization of an amplified template tSZ power spectrum). In both figures, we compare the results using various summary statistics: multifrequency power spectra (described in Sec.~\ref{sec.method_multifrequency_power_spec}), harmonic ILC (HILC) power spectra (with a few variants based on weight determination and parameter dependence determination that are described in Sec.~\ref{sec.method_hilc}), and NILC power spectra (described in Sec.~\ref{sec.method_nilc}). For LFI, we use 30000 simulations for each pipeline. For the Gaussian likelihood, we use 2000 simulations and show the distribution of maximum-likelihood estimates from each simulation. }
    \label{fig:final_posteriors_gaussiantsz}
\end{figure}

\begin{figure}[htb]
    \centering
    {\includegraphics[width=0.4\textwidth]{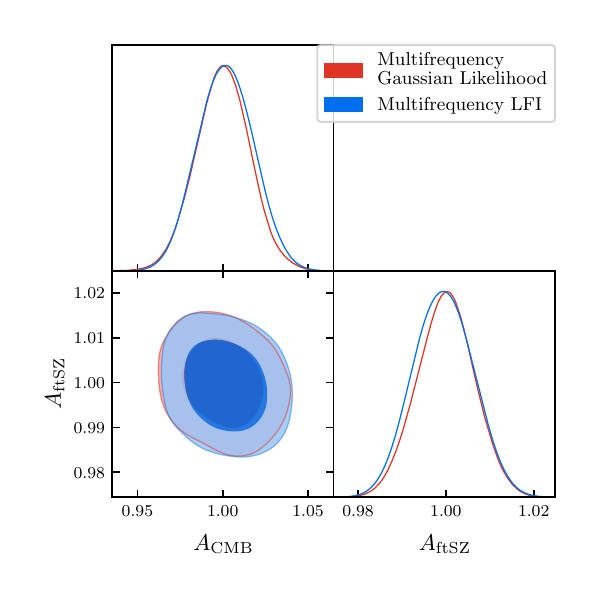}}
    {\includegraphics[width=0.4\textwidth]{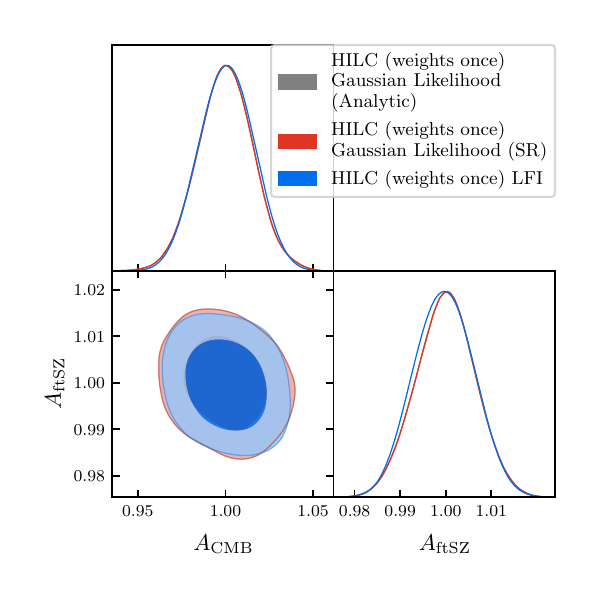}}
    {\includegraphics[width=0.4\textwidth]{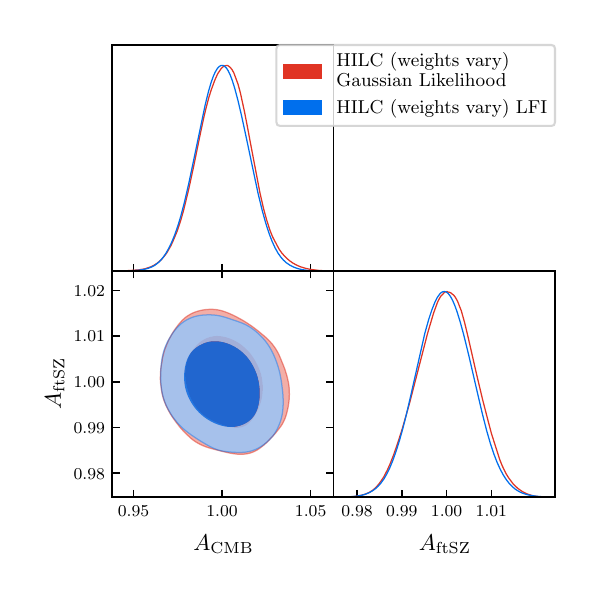}}
    {\includegraphics[width=0.4\textwidth]{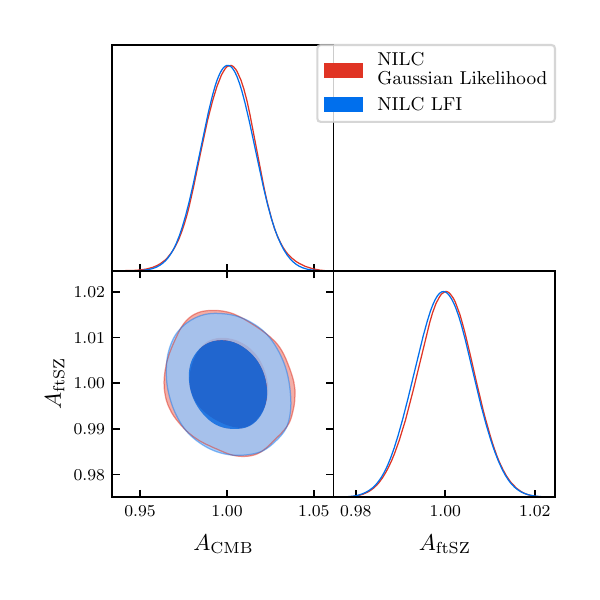}}
    \caption{Comparison of posteriors from LFI to those from a Gaussian likelihood for all the summary statistics considered in this work, computed on simulations consisting only of Gaussian random fields. The Gaussian likelihood is a reasonable assumption when every field in the simulation is a Gaussian random field.}
    \label{fig:lfi_vs_gaussianlkl_gaussiantsz}
\end{figure}

The posterior distributions for both LFI and the Gaussian likelihood (with posteriors computed via MLE here) are shown in Fig.~\ref{fig:final_posteriors_gaussiantsz}. From these plots, we see that multifrequency, HILC, and NILC power spectra yield nearly identical posteriors. Fig.~\ref{fig:lfi_vs_gaussianlkl_gaussiantsz} compares the posteriors from LFI and the Gaussian likelihood for each of the pipelines, validating the LFI procedure by showing that it recovers the same results as the Gaussian likelihood in this situation where we know that the Gaussian likelihood should hold. Table \ref{table:posteriors_gaussiantsz} shows the 1D marginalized posteriors for each of the methods discussed in this work, evaluated on this set-up of only Gaussian random fields.

\begin{table}[htb]
\setlength{\tabcolsep}{10pt}
\renewcommand{\arraystretch}{1.3}
    \begin{tabular}{|l|l|l|}
        \hline
         \textbf{Pipeline (using only Gaussian random fields)} &  $A_{\mathrm{CMB}}$ & $ A_{\mathrm{ftSZ}}$  \\
         \hline \hline 
         \textbf{Multifrequency PS} &MLE: $1.000 \pm 0.016 $    &MLE: $0.99998 \pm 0.0064$   \\
          -Gaussian likelihood &Fisher: \_\_ $\pm 0.016$    &Fisher: \_\_ $\pm 0.0065$   \\
           -analytic parameter dependence &MCMC: $0.999\pm 0.015$    &MCMC: $0.99997\pm 0.0064$   \\
           
         \hline
          \textbf{Multifrequency PS} &$1.002 \pm 0.016$  &$0.9994 \pm 0.0068$  \\
          -likelihood-free inference &  &  \\
        \hline \hline

        \textbf{Harmonic ILC} &MLE: $1.000 \pm 0.016$    &MLE: $0.99998 \pm 0.0064$   \\
        -fixed weights &Fisher: \_\_ $\pm 0.016$    &Fisher: \_\_ $\pm 0.0065$   \\
        -Gaussian likelihood &MCMC: $0.999\pm 0.016$    &MCMC: $1.0001\pm 0.0064$   \\
        -analytic parameter dependence &    &   \\
        \hline

        \textbf{Harmonic ILC} &MLE: $1.000 \pm 0.016$  &MLE: $1.0000 \pm 0.0064$   \\
        -fixed weights &Fisher: \_\_ $\pm 0.016$     &Fisher: \_\_ $\pm 0.0065$   \\
        -Gaussian likelihood &MCMC: $ 0.9997\pm 0.016$   &MCMC: $1.0006\pm 0.0062$   \\
        -parameter dependence from symbolic regression &   &   \\
        \hline
        
        \textbf{Harmonic ILC} &$1.001 \pm 0.016$    &$0.9996 \pm 0.0065$   \\
        -fixed weights &    &   \\
        -likelihood-free inference &    &   \\
        \hline

        \textbf{Harmonic ILC} &MLE: $1.001 \pm 0.015$    &MLE: $0.99995 \pm 0.0064$   \\
        -varying weights &Fisher: \_\_ $\pm 0.015$    &Fisher: \_\_ $\pm 0.0064$   \\
        -Gaussian likelihood &MCMC: $1.002\pm 0.015$    &MCMC: $0.9995\pm 0.0064$   \\
        -parameter dependence from symbolic regression &    &   \\
        \hline
        
        \textbf{Harmonic ILC} &$1.000 \pm 0.014$    &$0.9995 \pm 0.0062$   \\
        -varying weights &    &   \\
        -likelihood-free inference &    &   \\
        \hline
        \hline

        \textbf{Needlet ILC} &MLE: $1.001 \pm 0.015$ &MLE: $1.0000 \pm 0.0064$   \\
        -Gaussian likelihood &Fisher: \_\_ $\pm 0.015$   &Fisher: \_\_ $\pm 0.0064$  \\
        -parameter dependence from symbolic regression &MCMC: $1.000\pm 0.016$    &MCMC: $1.0002\pm 0.0064$   \\
        \hline
        
        \textbf{Needlet ILC} &$1.001 \pm 0.015$    &$0.9997 \pm 0.0065$   \\
        -likelihood-free inference &  &   \\
        \hline
        
    \end{tabular}
    \caption{Parameter constraints for all the methods considered in this work, as applied to simulations containing only Gaussian random fields (here the ``ftSZ" field is a Gaussian realization of an amplified tSZ power spectrum template). See Sec.~\ref{sec.method_multifrequency_power_spec}, \ref{sec.method_hilc}, and \ref{sec.method_nilc} for information on using multifrequency power spectra, harmonic ILC power spectra, and needlet ILC power spectra as summary statistics, respectively. We present 1D marginalized posterior values obtained with the different summary statistics using both likelihood-free inference (with 30000 simulations) and a Gaussian likelihood (with 2000 simulations). For the Gaussian likelihood, we compute posteriors using maximum-likelihood estimation (MLE), Fisher matrix calculation, and a Markov chain Monte Carlo (MCMC) algorithm. Note that we do not quote a central value for the Fisher matrix calculation since it only predicts the error bars (which are assumed to be symmetric in the Fisher calculation).}
    \label{table:posteriors_gaussiantsz}
\end{table}

\section{Effects of Varying Frequency Channels}
\label{app:3freqs}

In Sec.~\ref{sec.results} we consider a toy model in which we use two frequency channels: 90 and 150 GHz. In this appendix, we consider various other combinations of frequency channels to forecast the impact of NILC on parameter constraints.  Our physical sky model is identical to that in Sec.~\ref{sec.results}, comprising the CMB and ftSZ fields, but we evaluate these at different sets of frequencies here.  In particular, we consider the following sets of channels (all in GHz): 90, 120; 90, 150; 90, 280; 90, 120, 150; 90, 150, 280; 90, 150, 220; 280, 353; and 280, 353, 400. The noise models and beams are the same in each frequency channel. We compute the results for both the Gaussian likelihood and likelihood-free inference using each summary statistic (multifrequency power spectra, HILC with fixed weights, HILC with varying weights, and NILC), for each set of frequencies. We then estimate the area of the resulting 2D posteriors by extracting the covariance matrix and assuming an elliptical posterior (a crude approximation but sufficient for the rough estimation performed here). As described in Ref.~\cite{Fisher_ellipse}, the ellipse parameters are then given by 

\begin{equation}
    a^2 = \frac{\sigma_x^2+\sigma_y^2}{2} + \sqrt{\frac{(\sigma_x^2-\sigma_y^2)^2}{4} + \sigma^2_{xy}} \; , 
\end{equation}

\begin{equation}
    b^2 = \frac{\sigma_x^2+\sigma_y^2}{2} - \sqrt{\frac{(\sigma_x^2-\sigma_y^2)^2}{4} + \sigma^2_{xy}} \; , 
\end{equation}
and the area of ellipse is then 
\begin{equation}
    A = \pi ab \Delta \chi^2 \; , 
\end{equation}
where $\Delta \chi^2 = 2.3$ for the 68\% confidence interval.

\begin{figure}[htb]
    \centering
    {\includegraphics[width=0.3\textwidth]{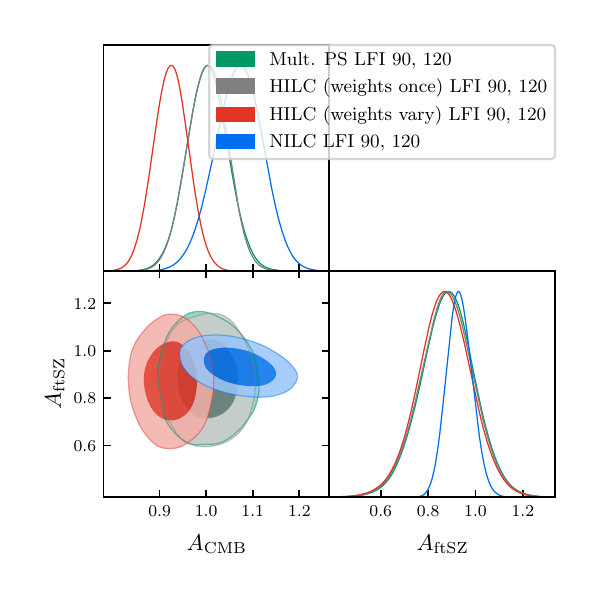}}
    {\includegraphics[width=0.3\textwidth]{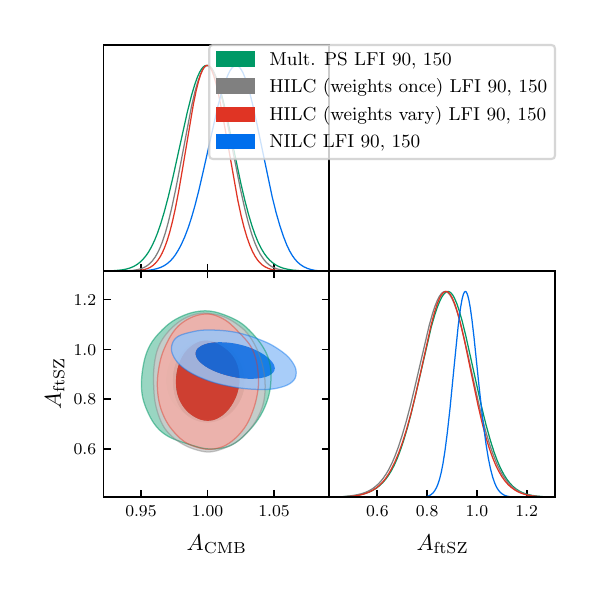}}
    {\includegraphics[width=0.3\textwidth]{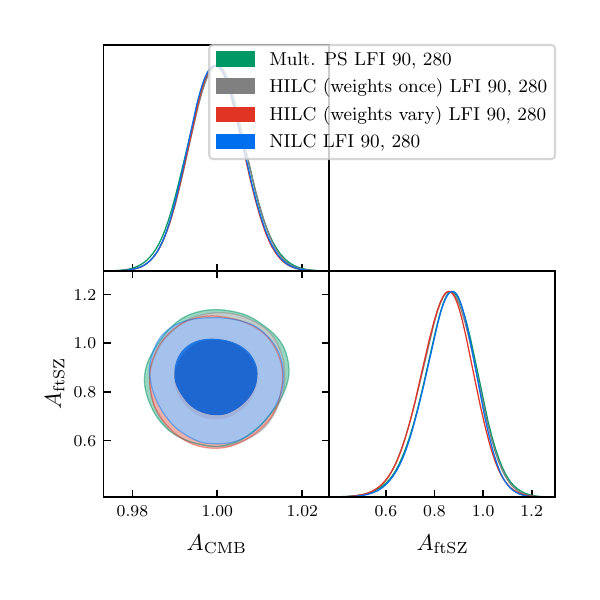}}
    {\includegraphics[width=0.3\textwidth]{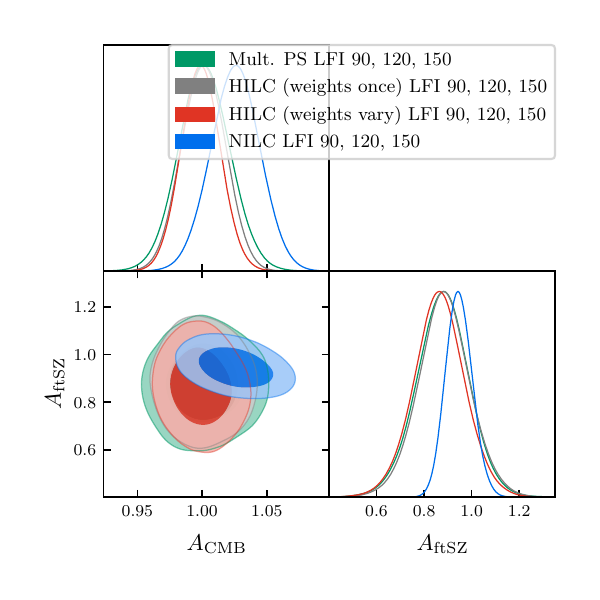}}
    {\includegraphics[width=0.3\textwidth]{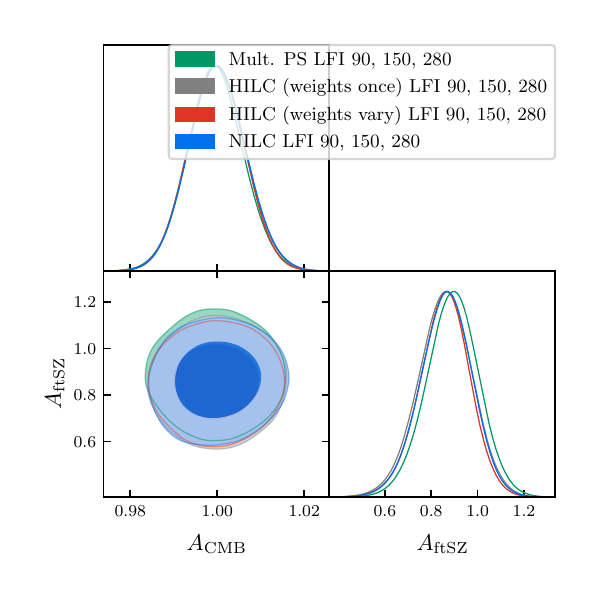}}
    {\includegraphics[width=0.3\textwidth]{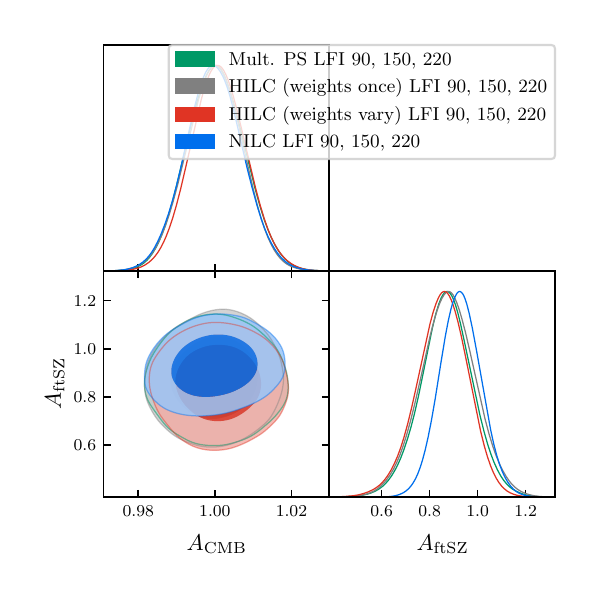}}
    {\includegraphics[width=0.3\textwidth]{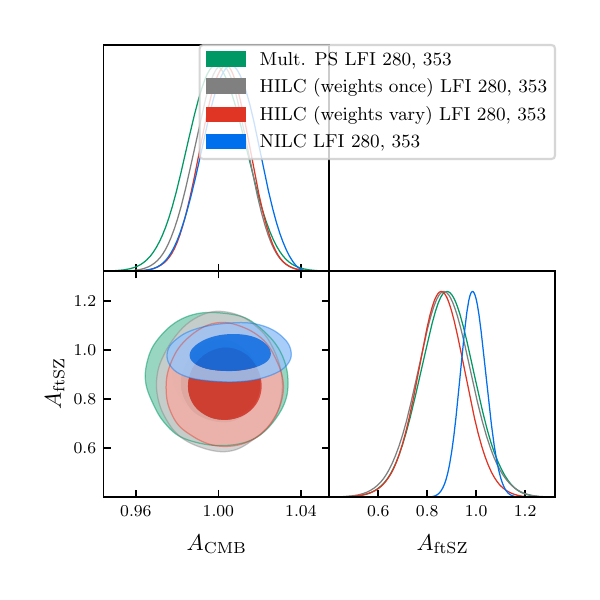}}
    {\includegraphics[width=0.3\textwidth]{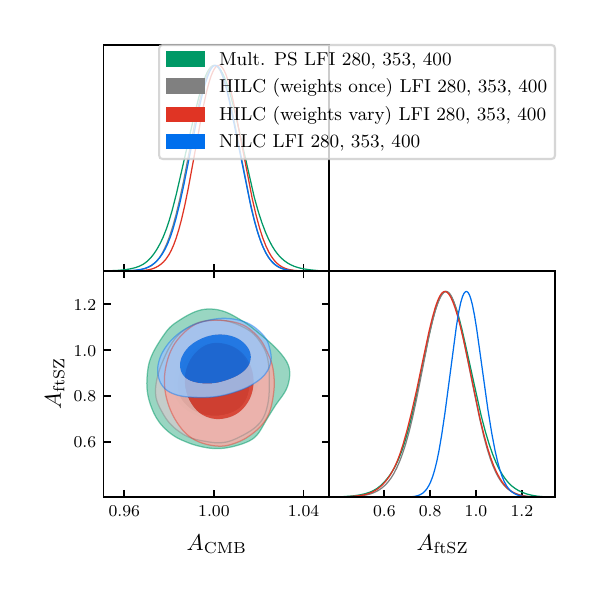}}
    \caption{Posteriors using various summary statistics and frequency combinations with LFI.}
    \label{fig:posteriors_varyfreqs_lfi}
\end{figure}

\begin{figure}[htb]
    \centering
    {\includegraphics[width=0.3\textwidth]{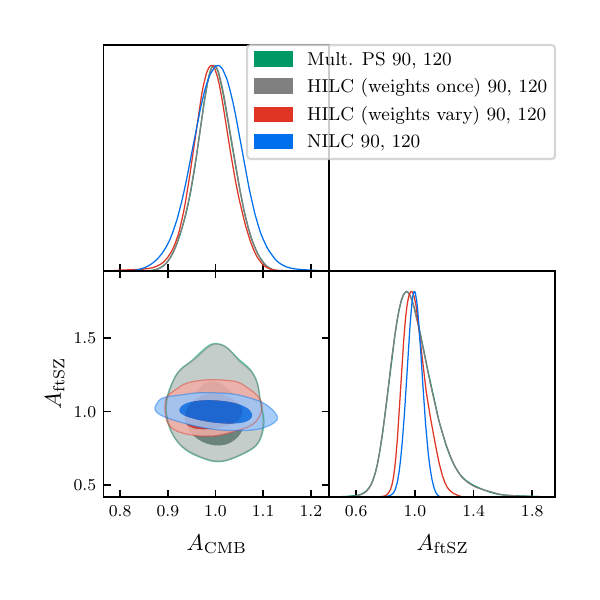}}
    {\includegraphics[width=0.3\textwidth]{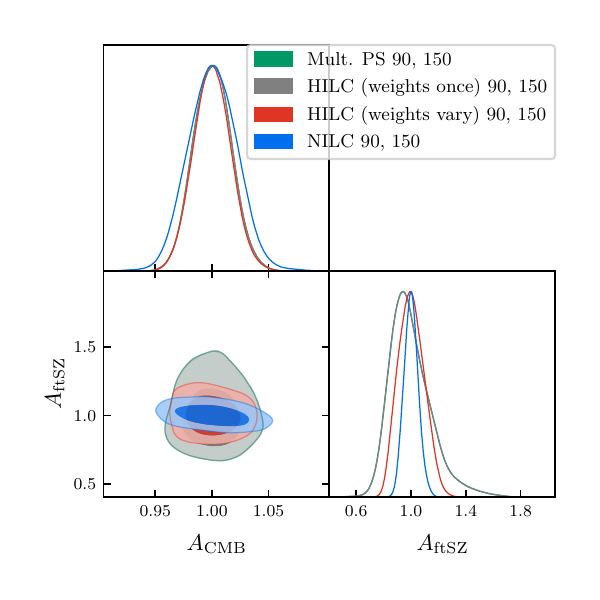}}
    {\includegraphics[width=0.3\textwidth]{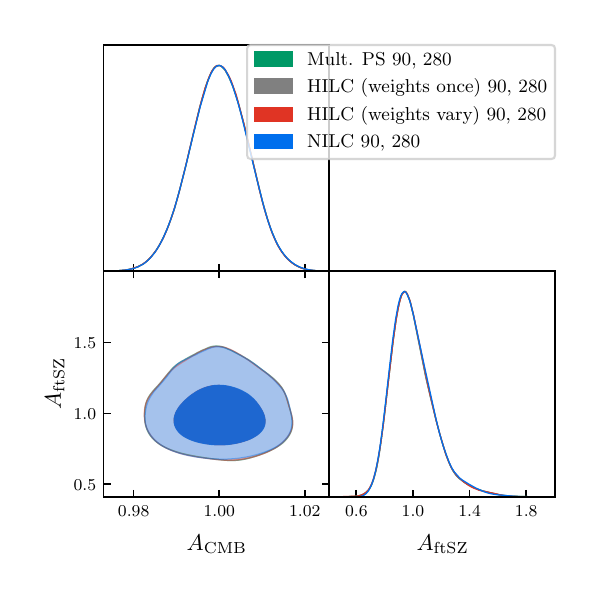}}
    {\includegraphics[width=0.3\textwidth]{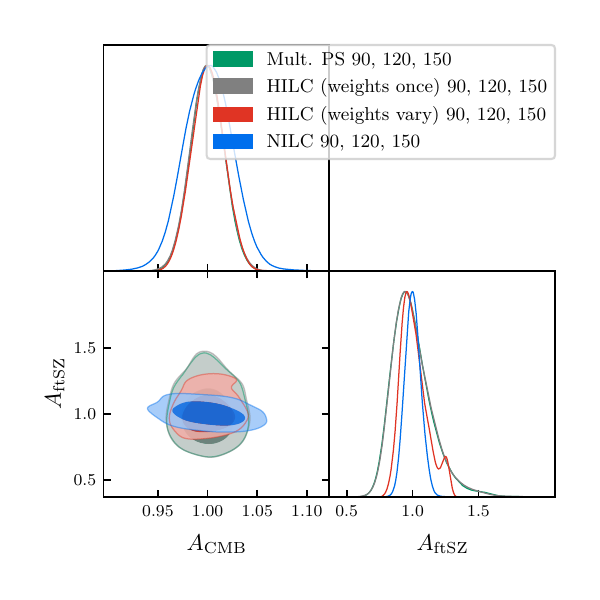}}
    {\includegraphics[width=0.3\textwidth]{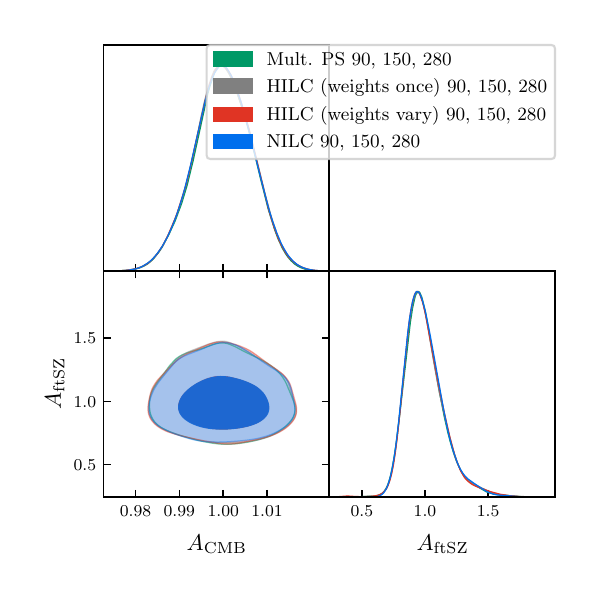}}
    {\includegraphics[width=0.3\textwidth]{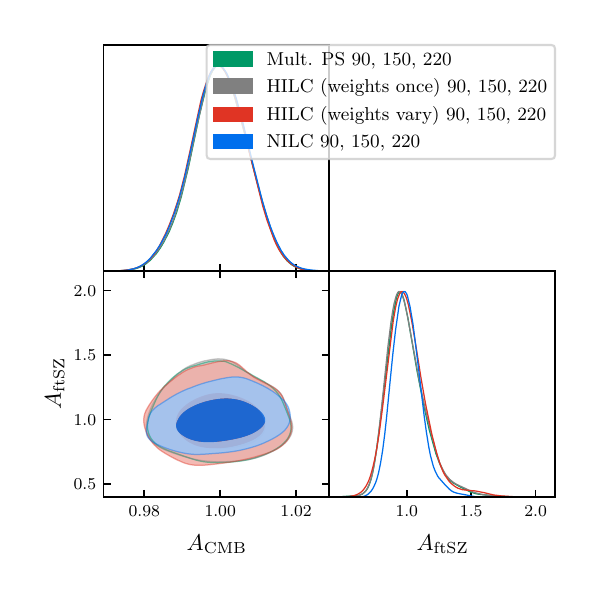}}
    {\includegraphics[width=0.3\textwidth]{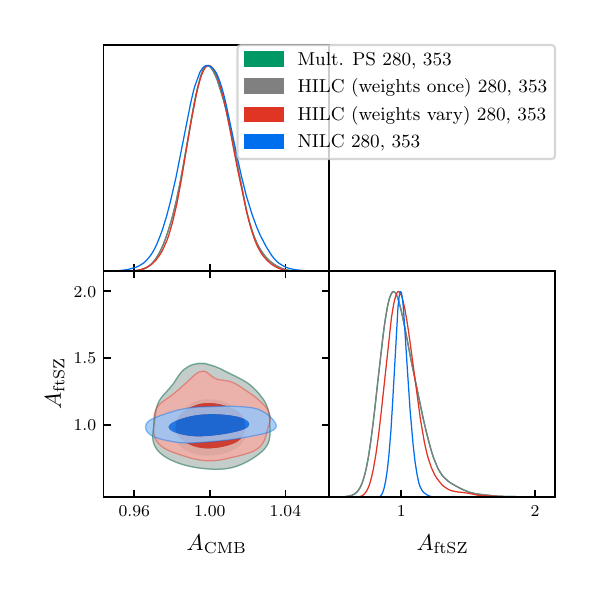}}
    {\includegraphics[width=0.3\textwidth]{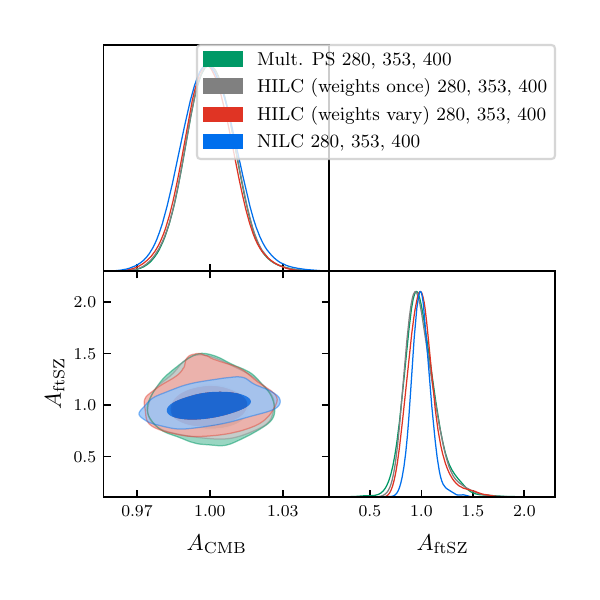}}
    \caption{Posteriors using various summary statistics and frequency combinations with Gaussian likelihoods. For the HILC (weights once) case, we use the analytic parameter dependence.}
    \label{fig:posteriors_varyfreqs_gaussianlkl}
\end{figure}

Table \ref{table:ellipse_area} details the 68\% confidence ellipse areas for each of the frequency sets for each summary statistic, using likelihood-free inference, and Fig.~\ref{fig:posteriors_varyfreqs_lfi} shows the posteriors, comparing the various summary statistics for each individual set of frequencies. Notably, the 2D posterior obtained using NILC power spectra as a summary statistic (blue contours in  Fig.~\ref{fig:posteriors_varyfreqs_lfi}) is the same or smaller than the 2D posteriors obtained using multifrequency power spectra as a summary statistic (green contours in  Fig.~\ref{fig:posteriors_varyfreqs_lfi}) for every set of frequencies. Table \ref{table:ellipse_area_gaussianlkl} and Fig.~\ref{fig:posteriors_varyfreqs_gaussianlkl} show the analogous information obtained using a Gaussian likelihood instead of LFI.

Using harmonic ILC with fixed weights, ``HILC (weights once)", often yields smaller 2D posterior areas than using multifrequency power spectra in the LFI case. From an information perspective, both methods contain the exact same information, as the former just rescales the latter by some fixed constants that are applied to every simulation in the same way. This indicates that the LFI may not yet have converged for the multifrequency power spectrum case. Thus, this simple rescaling into HILC power spectra may aid in the algorithm's ability to learn parameters by transforming the data into a space that is more immediately connected to the parameters of interest.

Also of note is that, in most cases, adding additional frequency channels increases the constraining power of a given summary statistic, where the constraining power is estimated by the area of the Fisher ellipse. As can be seen in Table~\ref{table:ellipse_area}, there are a few exceptions to this trend. From an information content standpoint, adding frequency channels can only add information and thus increase constraining power. Thus, we attribute minor deviations from the expected trend to lack of full LFI convergence (or need for more simulations in the Gaussian likelihood case) and the crude approximation that the 2D posteriors are elliptical.

\begin{figure}[htb]
    \centering
    {\includegraphics[width=0.325\textwidth]{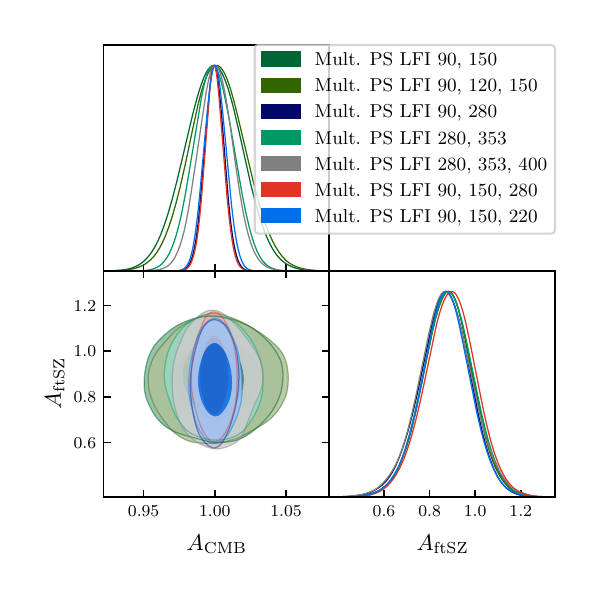}}
    {\includegraphics[width=0.325\textwidth]{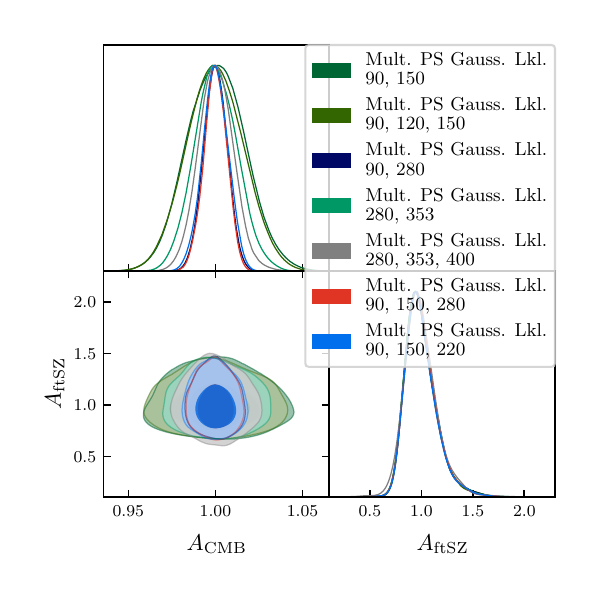}}
    {\includegraphics[width=0.325\textwidth]{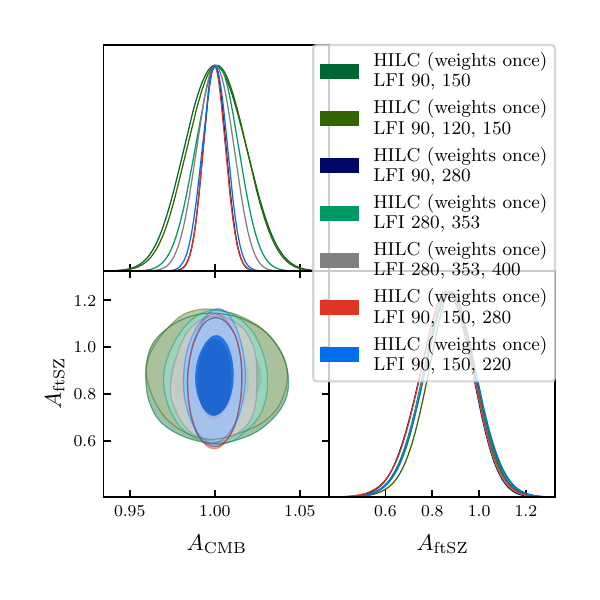}}
    {\includegraphics[width=0.325\textwidth]{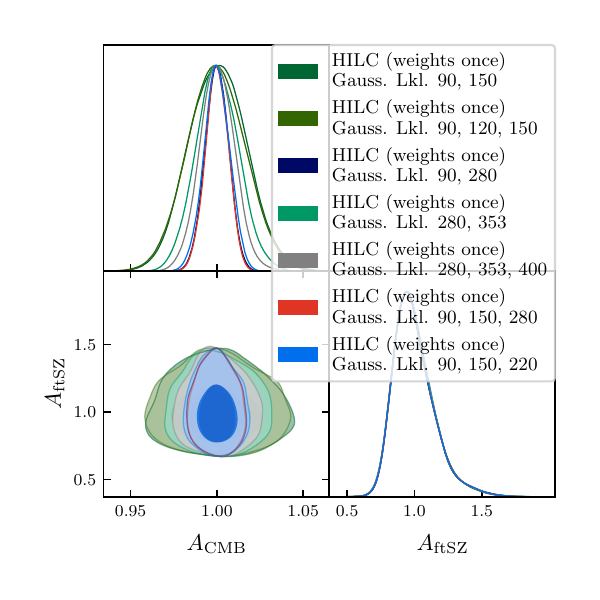}}
    {\includegraphics[width=0.325\textwidth]{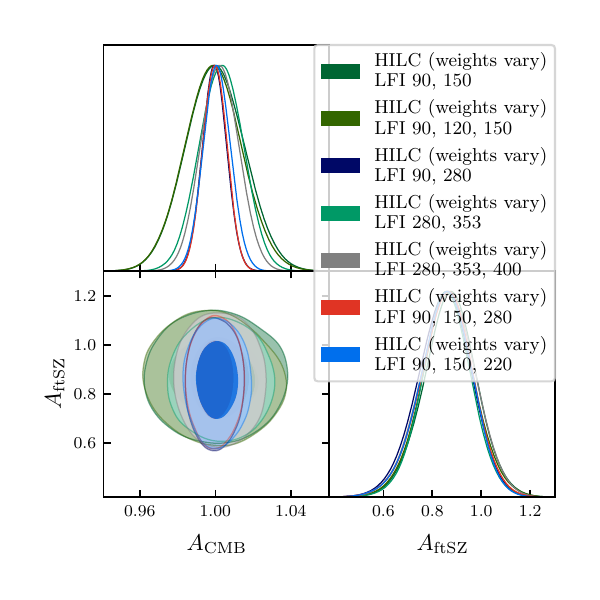}}
    {\includegraphics[width=0.325\textwidth]{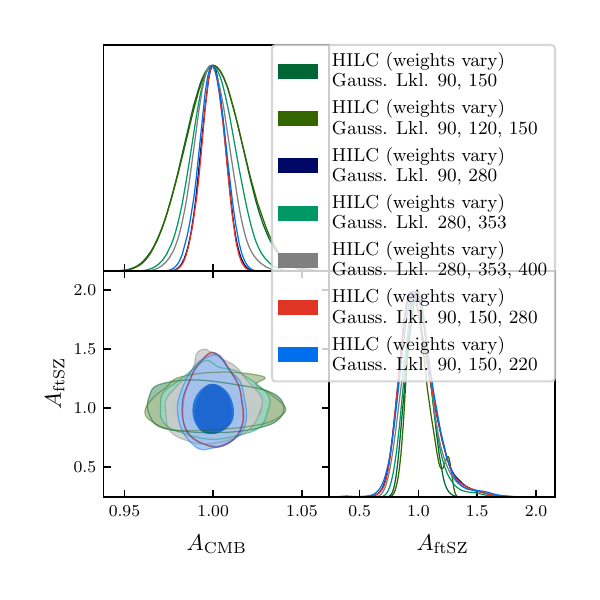}}
    {\includegraphics[width=0.325\textwidth]{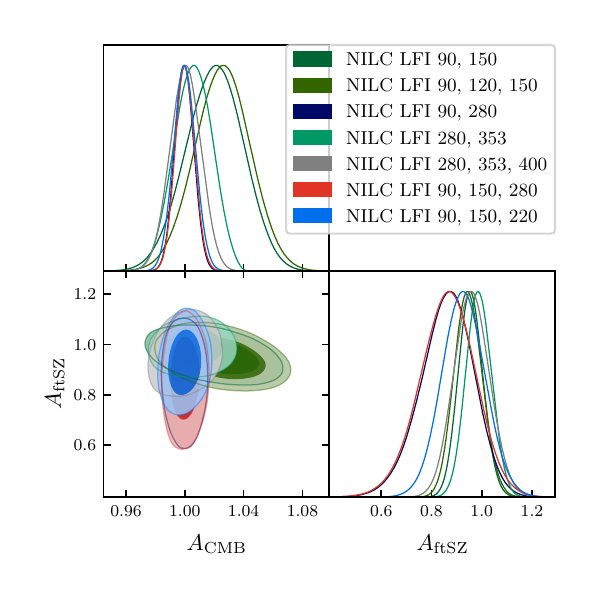}}
    {\includegraphics[width=0.325\textwidth]{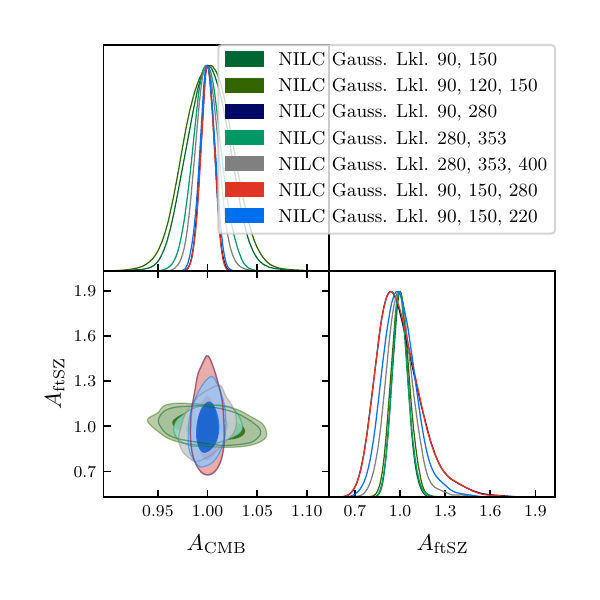}}
    \caption{Comparison of different frequency combinations used for each individual summary statistic, with both LFI and a Gaussian likelihood. For the HILC (weights once) Gaussian likelihood, we use the analytic parameter dependence.}
    \label{fig:posteriors_varyfreqs}
\end{figure}

In our set-up, we have selected the amplification factor of the Compton-$y$ field specifically such that the resulting ``ftSZ" power spectrum would have comparable magnitude to that of the CMB. Thus, depending on the frequency used, the tSZ SED could make the ftSZ power spectrum either larger or smaller than that of the CMB. This would play a role in which parameter is most well-constrained. We observe this effect in the NILC posteriors in Fig.~\ref{fig:posteriors_varyfreqs}, where the orientation of the posterior interestingly flips, for example, when switching from using 90 and 150 GHz channels to using 90, 150, and 280 GHz channels.

\clearpage

\begin{table}[htb]
\setlength{\tabcolsep}{10pt}
\renewcommand{\arraystretch}{1.3}
    \begin{tabular}{|l||l|l|l|l|}
    \hline
    \textbf{Frequencies (GHz)}&\textbf{Multifrequency PS} & \textbf{HILC (weights once)} &\textbf{HILC (weights vary)} &\textbf{NILC} \\
    \hline \hline
    90, 120& 36.0& 35.2& 30.6& 18.8 \\ \hline 
    90, 150& 16.0& 14.2& 12.2& 6.3 \\ \hline 
    90, 280& 5.5& 5.3& 5.1& 5.0\\ \hline
    90, 120, 150& 16.0& 13.4& 11.9 &7.2 \\ \hline  
    90, 150, 280& 5.3& 5.4& 5.0& 5.3 \\ \hline 
    90, 150, 220& 6.0& 6.1& 5.7& 4.6 \\ \hline 
    280, 353& 11.4& 10.3& 8.5& 4.6 \\ \hline 
    280, 353, 400& 10.8& 8.3& 8.2& 5.3 \\ \hline 
    \end{tabular}
    \caption{Area of 2D posteriors multiplied by a factor of 1000 for clarity (assuming elliptical posteriors) for various frequency combinations, using likelihood-free inference. We note that the elliptical approximation is crude in some cases.}
    \label{table:ellipse_area}
\end{table}

\begin{table}[htb]
\setlength{\tabcolsep}{10pt}
\renewcommand{\arraystretch}{1.3}
    \begin{tabular}{|l||l|l|l|l|}
    \hline
    \textbf{Frequencies (GHz)}&\textbf{Multifrequency PS} & \textbf{HILC (weights once)} &\textbf{HILC (weights vary)} &\textbf{NILC} \\
    \hline \hline
    90, 120& 46.8& 46.5& 25.4& 19.1 \\ \hline 
    90, 150& 19.5& 19.5& 11.2& 8.3 \\ \hline 
    90, 280& 7.9& 7.9& 7.8& 7.9 \\ \hline
    90, 120, 150& 19.1& 19.4& 12.4 &10.2 \\ \hline  
    90, 150, 280& 7.5& 7.7& 7.8& 7.7 \\ \hline 
    90, 150, 220& 8.5& 8.7& 9.0& 6.6 \\ \hline 
    280, 353& 14.4& 14.4& 12.3& 5.8 \\ \hline 
    280, 353, 400& 14.2& 11.8& 11.7& 7.6 \\ \hline 
    \end{tabular}
    \caption{Area of 2D posteriors multiplied by a factor of 1000 for clarity (assuming elliptical posteriors) for various frequency combinations, using Gaussian likelihoods. For the HILC (weights once) case, we use the analytic parameter dependence. We note that the elliptical approximation is crude in some cases. In particular, for the case of 280, 353, and 400 GHz, it appears that the HILC (weights once) posterior is significantly smaller than the multifrequency PS posterior. However, the 1D marginalized posteriors are $A_{\mathrm{CMB}}=1.000\pm 0.010$ and $A_{\mathrm{ftSZ}}=1.00^{+0.10}_{-0.18}$ in the HILC (weights once) pipeline, and $A_{\mathrm{CMB}}=1.000\pm 0.010$ and $A_{\mathrm{ftSZ}}=0.999^{+0.12}_{-0.17}$ in the multifrequency PS case, and thus, the two methods actually produce nearly identical numerical posteriors.  }
    \label{table:ellipse_area_gaussianlkl}
\end{table}

\end{appendices}

\clearpage
\bibliography{refs}
\end{document}